\documentclass[lettersize,journal]{IEEEtran}

\usepackage{subfigure}
\usepackage{dsfont}
\usepackage{graphicx}
\usepackage{amsmath}

\usepackage{amssymb}
\usepackage{bm}
\usepackage{float}
\usepackage{booktabs}
\usepackage{multirow}
\usepackage{makecell} 
\usepackage{verbatim}

\usepackage{color}
\usepackage{soul}
\soulregister\cite7
\soulregister\ref7
\makeatletter
\newcount\SOUL@minus 
\makeatother

\renewcommand{\hl}[1]{#1} 

\begin{document}
\bstctlcite{MyBSTcontrol}

\title{TexSenseGAN: A User-Guided System for Optimizing Texture-Related Vibrotactile Feedback Using Generative Adversarial Network}

\author{Mingxin Zhang, Shun Terui, Yasutoshi Makino, Hiroyuki Shinoda
\thanks{This work was supported in part by the JST SPRING JPMJSP2108, CAO-NEDO SIP 23201554-0, and JST CREST JPMJCR18A2.}
\thanks{Mingxin Zhang, Shun Terui, Yasutoshi Makino, Hiroyuki Shinoda are with the Department of Complexity Science and Engineering, The University of Tokyo, Chiba 277-8561, Japan (e-mail: m.zhang@hapis.k.u-tokyo.ac.jp, terui@hapis.k.u-tokyo.ac.jp, yasutoshi\_makino@k.u-tokyo.ac.jp, hiroyuki\_shinoda@k.u-tokyo.ac.jp).}}

\maketitle

\begin{abstract}

Vibration rendering is essential for creating realistic tactile experiences in human-virtual object interactions, such as in video game controllers and VR devices. By dynamically adjusting vibration parameters based on user actions, these systems can convey spatial features and contribute to texture representation. However, generating arbitrary vibrations to replicate real-world material textures is challenging due to the large parameter space. This study proposes a human-in-the-loop vibration generation model based on user preferences. To enable users to easily control the generation of vibration samples with large parameter spaces, we introduced an optimization model based on Differential Subspace Search (DSS) and Generative Adversarial Network (GAN). With DSS, users can employ a one-dimensional slider to easily modify the high-dimensional latent space to ensure that the GAN can generate desired vibrations. We trained the generative model using an open dataset of tactile vibration data and selected five types of vibrations as target samples for the generation experiment. Extensive user experiments were conducted using the generated and real samples. The results indicated that our system could generate distinguishable samples that matched the target characteristics. Moreover, we established a correlation between subjects' ability to distinguish real samples and their ability to distinguish generated samples.
\end{abstract}


\section{Introduction}
\begin{figure*}
  \includegraphics[width=\textwidth]{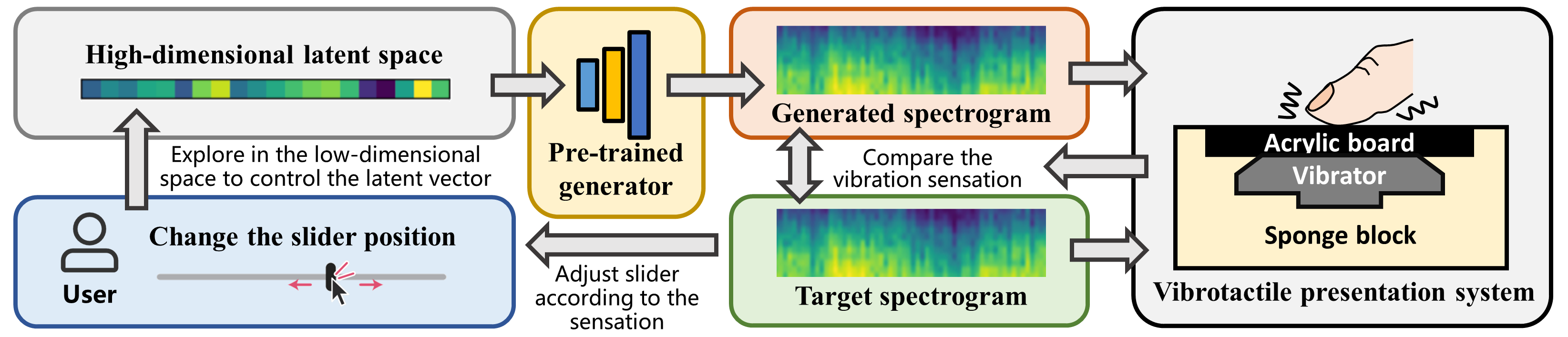}
  \caption{The structure of our optimization system. Users can easily control the high-dimensional latent vector with a 1-D slider, and the generator can give spectrograms according to the latent vector. Users can compare the tactile sensation of the generated vibration and the real target, and update the optimizer by the slider position corresponding to the closest result to the target.}
  \label{fig:system}
\end{figure*}

Virtual realm built by information technologies has become an integral part of our daily life. Among various interaction technologies, haptics, which enables natural and intuitive tactile interaction with virtual objects and surfaces, allows users to engage with this information-rich virtual world and improves the user experience \cite{10.1145/2642918.2647407, tong2023survey}. In the research field of haptics, the rendering of texture of objects has been an important topic to augment the user experience in virtual reality (VR) or other application scenarios \cite{Incorporating-the-Perception, GAN-based-image-to-friction, Midair-Haptic-Optic-Display}. Texture stimuli are characterized by spatially and temporally correlated signals. Many studies have modeled this kind of stimuli using parameters like velocity, direction, and force, often employing approaches such as auto-regression models \cite{10160503}. Furthermore, research has shown that realistic tactile feedback can be achieved even using only temporal stimuli, such as vibrations \cite{10198337, konyo2005tactile}. In this study, we propose a new framework for generating arbitrary vibration waveforms with the aim of presenting texture related feedback.

Currently, common vibration presenting methods include playing back vibrations recorded using sensors or microphones, and manually designing vibration waveforms through tools like equalizers. While manual design is valuable in scenarios where direct measurement is unavailable, such as video games or when enhancing specific tactile sensations, it is limited by the complexity of simultaneously adjusting multiple parameters \cite{Incorporating-the-Perception}, and the diversity and abundance of tactile textures in the physical world can also make this kind of adjustment impractical to create specific models for each texture.



To address these challenges, studies have explored reconstructing vibrations using variety methods like generative models \cite{GAN-based-image-to-friction, Preference-Driven-Texture-Modeling}, optimizing objective functions and delivering effective results. However, human subjective perception often diverges from mathematical metrics or direct parameter adjustments, making it beneficial to let users guide the generative process for tasks involving subjective evaluation. This human-in-the-loop approach incorporates human judgment and preferences into optimization, enabling models to produce results that align with individual perceptions \cite{Preference-Driven-Texture-Modeling, Sequential-Line-Search, Differential-Subspace-Search}. These mentioned points highlight the importance of developing a system capable of intuitively generating vibrations and dynamically accommodating a wide range of vibration patterns.



As mentioned above, adjustment of vibrotactile signals with multiple parameters manually is a challenging task. Techniques like Sequential Line Search (SLS) \cite{Sequential-Line-Search} allow users control multiple parameters via a 1-D slider simultaneously, making the process more practical. However, this kind of conventional Bayesian optimization approaches may struggle with complexities, especially in scenarios where we want to use deep learning models with a better generalization performance.

In this study, we introduced TexSenseGAN, a user-driven system combining Generative Adversarial Networks (GAN) and Differential Subspace Search (DSS) to optimize vibration stimuli for subjective ``texture-related sense" generation (Fig. \ref{fig:system}). A simple 1-D slider interface was developed to explore the high-dimensional tactile latent space easily. Emphasizing human perception, we integrated subjective preferences into optimizing and evaluating tactile feedback. The experimental results showed a 58\% accuracy rate for distinguishing generated vibrations in a five-class task, compared to 74\% for real vibrations. A linear regression analysis revealed a correlation between these accuracy rates, demonstrating the system’s ability to recreate distinct vibration characteristics guided by user input. Despite extensive research on human-in-the-loop approaches in various interaction methods, their application in haptics remains underexplored. We believe this work will contribute to the development of more intuitive and realistic haptic feedback systems.

\section{Related Work}
\subsection{Vibration Tactile Presentation}
Vibrotactile haptic devices have become a key focus in tactile presentation systems. Several studies aim to replicate various sensory experiences. For example, piezo actuators are used to simulate physical button sensations on touchscreens, as demonstrated by Sadia et al. \cite{sadia2020data}. Additionally, wearable devices, such as FingerX \cite{tsai2022fingerx}, enhance the sense of object shapes in VR environments.

Among the array of sensory modalities, the recreation of tactile textures has garnered significant attention. Through the capture of sound recordings while traversing real surfaces, the corresponding texture-specific vibrations can be reproduced via audio signals \cite{10198337}. Nevertheless, note that this research underscores the necessity of meeting specific requirements in terms of both time resolution and frequency resolution to effectively replicate tactile sensations. Furthermore, it highlights the critical role of parameter design in shaping the characteristics of the vibrotactile signal.

\subsection{Generative Neural Network}
In addressing parameter design challenges in the domain of vibrotactile texture reproduction, a variety of optimization algorithms have been applied, as highlighted in recent works \cite{park2021vibrotactile, chan2022investigating}. However, for scenarios involving a substantial number of parameters, deep learning techniques, which recently have seen significant advancements, have emerged as a more effective solution for managing the intricate optimization aspects of texture generation.

Autoencoders have been used in some studies for compressing and reconstructing tactile signals owing to their straightforward architecture and unsupervised training approach \cite{liu2023online, li2021autoencoder}. Nevertheless, the compression process can introduce oversmoothing, resulting in the loss of fine-grained details, making autoencoders difficult to reproduce detailed texture related vibration signals. Although increasing the dimension of the compressed latent vector can solve the problem, it makes operations, such as optimization, difficult. Therefore, the use of generative models, rather than a decoder within the autoencoder framework, to reconstruct vibration signals has become a prudent choice.

Generative Adversarial Networks, a kind of notable generative model, are widely recognized for their success in image generation and are increasingly applied to tactile reconstruction tasks \cite{Preference-Driven-Texture-Modeling, GAN-based-image-to-friction, TactGAN}. A GAN consists of two main components: the generator and discriminator. The generator produces samples from a specified distribution, often Gaussian noise, while the discriminator's role is to distinguish between generated samples and real ones. Through iterative training, the discriminator's ability to differentiate between genuine and generated samples improves, which in turn enhances the generator's capability to produce increasingly realistic samples.

The remarkable generative capabilities of GAN enable it to generate samples from random noise as well as from specific distributions, affording controlled sample generation opportunities. For example, prior research has explored the generation of surface images from vibrations \cite{cai2021visual}, and conversely, the generation of vibration signals from texture images \cite{GAN-based-image-to-friction, TactGAN}. This characteristic of GAN, which encompasses reconstruction as well as the generation of ``new'' samples conforming to a particular distribution while introducing variations, positions GAN as particularly well-suited for optimization problems in which we seek optimal solutions through continuous exploration of the latent space.

\subsection{Human-in-the-loop Optimization}
While GANs excel at generating samples that mathematically resemble real data, in some human evaluation conditions they do not inherently account for subjective cognition. In particular, we need to determine whether a small loss value equates to a similar perception or cognition. Basic computational metrics can highlight statistical differences between generated and real samples but fail to capture human subjective perception of quality. Research has been conducted to improve GAN architectures and use better loss functions to track training progress, reducing the need for developers to stare at generated samples to figure out failure modes \cite{arjovsky2017wasserstein}. However, the study acknowledges that they do not provide a full quantitative evaluation of generative models. More broadly, the challenge of quantitatively evaluating GAN performance is still an open problem, with no standard metrics agreed upon \cite{borji2019pros}. This highlights a gap between statistical measures of ``realness" and human subjective judgment, especially in tasks like image and audio generation where human perception still plays a key role in evaluation. Consequently, various studies have focused on optimization models incorporating a human-in-the-loop mode, which integrates human involvement into the iterative optimization process. This approach utilizes human expertise to guide the optimization direction.

Some human-in-the-loop systems focus solely on direct parameter optimization \cite{fang2022human} of several parameter dimension, while other explored optimization algorithms, such as Bayesian optimization \cite{Sequential-Line-Search, chan2022investigating}, to enable simultaneous control of multiple parameters. 
Furthermore, to overcome challenges when dealing with relatively large latent spaces, there have been some studies applying deep learning models in the optimization process to increase the capability of parameter dimension \cite{Preference-Driven-Texture-Modeling, Differential-Subspace-Search}. Another significant challenge in the optimization of deep learning models originates from the difficulty of incorporating human decision-making throughout several training epochs \cite{Preference-Driven-Texture-Modeling}.

A previous research offers users the choice to select the optimal texture for optimizing the latent vector instead of optimizing the neural network itself \cite{Preference-Driven-Texture-Modeling}. This approach provides an effective solution for optimizing deep learning models. In our own research, we adopted a similar strategy, employing pre-trained deep learning models and seeking positions in the latent space that correspond to optimal outputs. Furthermore, we employed continuous exploration of the latent space through sliders rather than discrete user interaction methods, facilitating a seamless and comprehensive exploration process.

\section{Methods}
\begin{figure*}[h]
  \centering
  \includegraphics[width=\linewidth]{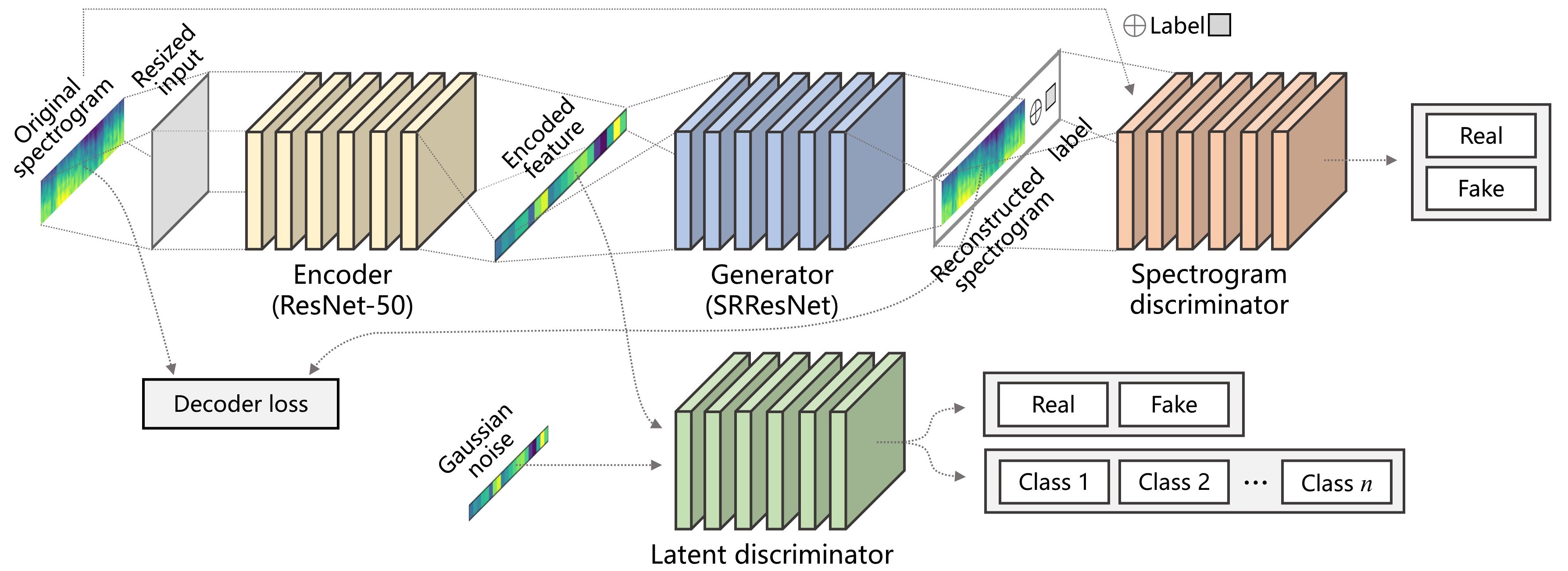}
  \caption{The structure of the GAN model in this research. The model can generate vibrotactile spectrograms from the latent space built by the ResNet-50 encoder.}
  \label{fig:ACAAE}
\end{figure*}

In this section, we will introduce the design and the implementation of our tactile vibration generation system. The whole human-in-the-loop optimization process is shown in Fig. \ref{fig:system}. Users can change the vibration and obtain the closest result to the target by changing the slider position. First, we will introduce the framework of the optimization method, and then introduce the design, structure and training of our generative model (as shown in Fig. \ref{fig:ACAAE}).

\subsection{Human-in-the-loop Differential Subspace Search}
To enable individuals to effectively leverage their knowledge for optimization purposes, it is important to ensure that users possess the ability to freely manipulate the resultant oscillatory patterns during each iteration of the optimization loop. However, exerting precise control over a multitude of parameters within a high-dimensional latent space is an intricate challenge. 

An efficacious solution to this quandary is presented by the DSS \cite{Differential-Subspace-Search}. DSS facilitates optimization in a manner that permits users to maintain command over the high-dimensional latent space while concurrently exploring a relatively lower-dimensional latent space. This approach streamlines the manipulation process for human users, rendering it more manageable and accessible.

The optimization problem can be represented as 
\begin{equation}
\bm{z}^* = \underset{\bm{z}\in{\bm{Z}}}{\mathrm{argmax}}\;E(G(\bm{z})).
\end{equation}
Here, the generator function $G$ assumes an input latent vector $\bm{z}$ from the high-dimensional latent space $\bm{Z}$, and the evaluation function $E$ quantifies the quality of the generated output. In our specific context, the choice of the function $E$ is contingent upon the subjective perception of the user. The objective of the optimization is to discover the optimal latent vector $\bm{z}^*$ that maximizes the value of the evaluation function $E$ through human-guided exploration within the latent space \cite{Differential-Subspace-Search}.

Consequently, the iterative update in the optimization process can be expressed as
\begin{equation}
\label{equ:iteration}
\bm{z}^{(k+1)} = \bm{z}^{(k)} + \alpha{(\frac{\partial E(\bm{z}^{(k)})}{\partial \bm{z}})^T}.
\end{equation}
The step size $\alpha$ is chosen to be greater than zero to ensure maximization of the value of the evaluation function.

To simplify user control over the latent vector, DSS introduces an operator denoted as $p_{\bm{z}^{(k)}}(w)$. This operator transforms a low-dimensional latent vector, specifically a single number $w$ that represents the slider position within the 1-D space $W$, into the corresponding position $\bm{z}^{(k)}$ within the original high-dimensional latent space $\bm{Z}$. This approach allows users to manipulate only the low-dimensional subspace. Consequently, Eq. \ref{equ:iteration} can be modified as follows:
\begin{equation}
\bm{z}^{(k+1)} = \bm{z}^{(k)} + p_{\bm{z}^{(k)}}(w).
\end{equation}
To derive the operator $p_{\bm{z}^{(k)}}(w)$, the determination of the subspace $W$ can be accomplished through singular value decomposition (SVD) applied to the Jacobian matrix of the generator \cite{Differential-Subspace-Search}.

\subsection{Generator Design}
We designed our vibrotactile generator architecture based on the framework of the super-resolution residual network (SRResNet) \cite{SRResNet, 10.1007/978-3-319-93399-3_3}. Our motivation was to effectively address high-frequency image details. There is an input convolutional layer, 16 residual blocks, a mid convolutional layer, a 4-times upscale module consisting of two convolutional layers and two pixel shuffle layer, and an output convolutional layer in the SRResNet. We introduced a Fully Connected (FC) layer to the input end of the network as the resize layer for dimension transformation, to ensure that our latent vector can be received by SRResNet. Additionally, we made an adjustment to the final convolutional layer of the network to tailor the output to the required size, specifically to generate the desired spectrogram.  

\begin{figure}[tb]
  \centering
  \includegraphics[width=\linewidth]{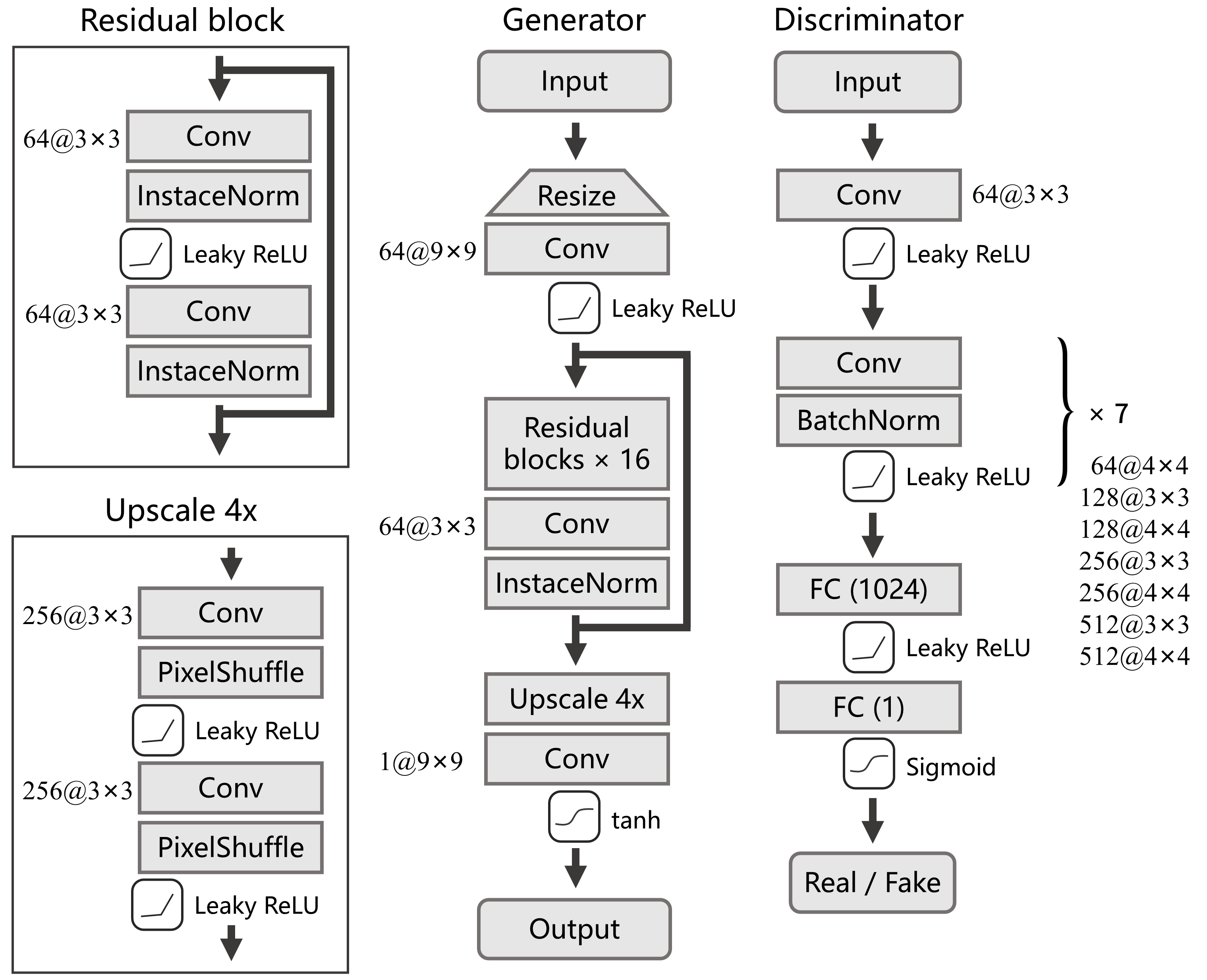}
  \caption{The structure of the SRResNet-based GAN.}
  \label{fig:SRGAN}
\end{figure}

To ensure the effectiveness of the generator, it is essential to employ an appropriate discriminator. A discriminator that performs exceptionally well or poor can hinder the training of the generator. In our study, we utilized a discriminator with a structure similar to the one presented in \cite{SRResNet}. This discriminator comprises eight convolutional layers, albeit with adjustments to the channel numbers to align it with the specific requirements of our task. The comprehensive architecture of the GAN is shown in Fig. \ref{fig:SRGAN}.

The generator and discriminator were trained simultaneously. In the training process, the generator $G$ learns how to generate new samples closer to the real sample, while the discriminator $D$ will discriminate the generated samples. The goal of the training can be represented as follows:
\begin{equation}
\begin{aligned}
\underset{G}{\mathrm{min}}\;\underset{D}{\mathrm{max}}\;&\mathds{E}_{x \sim p_{data}(x)}[\log(D(x))]\\
+&\mathds{E}_{z \sim p(z)}[\log(1-D(G(z))],
\end{aligned}
\end{equation}
where $z$ represents a random noise from the distribution $p(z)$ (Gaussian distribution in this research), and the $p_{data}(x)$ denotes the distribution of the real data $x$. This competitive process encourages the generator to generate samples that are indistinguishable from real data.

\subsection{Generation with Conditional Knowledge}
\label{sec:generator}

While GAN has outstanding generation performance, the training process is unstable. Additionally, GAN generate samples from random noise, resulting in uncontrollable outcomes. However, our objective is to exert control over the latent space, enabling the model to generate spectrograms that closely resemble real samples belonging to specific classes. To address these challenges, we built our model combining the advantages of GANs and autoencoders, and utilized conditional information. The inspiration for our model draws from previous works, such as Auxiliary Classifier GAN (ACGAN) \cite{odena2017conditional}, Autoencoding GAN (AEGAN) \cite{lazarou2020autoencoding}, and Conditional Adversarial Autoencoder (CAAE) \cite{Conditional-Adversarial-Autoencoder} to enhance the training process and enable the model to generate samples belonging to particular classes.

The model shown in Fig. \ref{fig:ACAAE} contains four parts: encoder, generator, spectrogram discriminator, and latent discriminator. The encoder extracts the feature vector from the original spectrogram, while the generator utilizes this feature vector to reconstruct the spectrogram. Similar to the Adversarial Autoencoder (AAE) paradigm, a latent discriminator is employed to encourage the encoder's output to conform to a predetermined prior distribution. This ensures that the latent space is evenly distributed according to the specified prior distribution, allowing the generator to effectively learn a mapping from this prior distribution to the spectrogram distribution, as proposed in \cite{makhzani2015adversarial}. Additionally, a determined distribution will benefit the following optimization of DSS. To make the feature containing richer information, we utilized the spectrogram discriminator as a classifier and applied the auxiliary classification loss, such as the ACGAN. It will work as the discriminator and classifier when receiving the encoded feature, and only as the discriminator when receiving the real gaussian noise.

To avoid data loss in compression and over-smoothing reconstruction, we used a generator together with the decoder loss, and use a weight value to balance them. The generator and spectrogram discriminator were structured according to the principles of CAAE. Notably, with the label as the conditional information, the generator can generate samples belonging to certain classes. The discriminator can learn the relationship in the pairs of the spectrogram and the corresponding class; thus, help the generator to provide correct spectrograms. The training goal can be described as follows:
\begin{equation}
\begin{aligned}
\underset{E,G,C}{\mathrm{min}}&\;\underset{D_z,D_s}{\mathrm{max}}\lambda \mathcal{L}(G(E(x),y),x)
+\gamma TV(G(E(x),y))\\
+&\mathds{E}_{z \sim p(z)}[\log(D_z(z))]\\
+&\mathds{E}_{x,y \sim p_{data}(x,y)}[\log(1-D_z(E(x)))]\\
+&\mathds{E}_{x,y \sim p_{data}(x,y)}[\sum_i y_i \log(C(E(x)))]\\
+&\mathds{E}_{x,y \sim p_{data}(x,y)}[\log(D_s(x,y))]\\
+&\mathds{E}_{x,y \sim p_{data}(x,y)}[\log(1-D_s(G(E(x)),y))],
\end{aligned}
\end{equation}
where $D_z$ represents the latent discriminator, $E$ and $G$ represent the encoder and generator, $D_{s}$ represents the spectrogram discriminator, and $C$ represent the auxiliary latent classifier, respectively. $x$ and $y$ represent the real data and the label. $TV$ represents the total variation loss. $\lambda$ and $\gamma$ are the weights for the decoder loss and the total variation loss, which can balance the smoothness and high resolution \cite{Conditional-Adversarial-Autoencoder}. In this study, the weight setting is $\lambda=100$ and $\gamma=10$.

We initialized the encoder using a pre-trained ResNet-50 model that was trained on the ImageNet1K\_V2 dataset, and introduced an FC layer to resize the spectrogram to the shape corresponding to the pre-trained model input. The generator and spectrogram discriminator follow the network structures illustrated in Fig. \ref{fig:SRGAN}, maintaining consistency with those architectures. The latent discriminator contains two FC layers: 1 output layer for the discrimination and 1 output layer for the classification. To apply the label as conditional information, we used one-hot labels and used an FC layer to resize the label vector to the dimension corresponding to the feature map. The label information will be added to the feature map as a channel.

\section{Experiments}
In this section, we will demonstrate the implementation details of the system and our experiment design.
\subsection{Data Preparation}
In this study, we employed the LMT108 dataset \cite{strese2016multimodal}, which includes recordings of 108 distinct textures along with their corresponding surface images. To construct our training dataset, we utilized the sound recording files capturing the motion of a texture explorer traversing various surfaces.

Seven groups (G1–G4, G6, G8, and G9) were selected from the total of nine groups in LMT108. According to a previous study, the original stimuli of G5 and G7 groups were found to evoke a tactile sensation that was similar to that of other groups \cite{10198337}, therefore these two groups were excluded from the study. While more data samples could enhance the generative model's learning, highly complex data can also hinder model convergence. Additionally, since we used an ACGAN-like auxiliary classifier to control the generation process, using the full LMT108 dataset would require the classifier to learn a challenging 108-class classification. This type of multi-classification problem represents a challenging task in deep learning field. Given that it is not the primary focus of our research, we chose to avoid scenarios involving an excessively large number of categories. Although CGAN could incorporate conditional information without the need for an auxiliary classifier, after comparing its generative performance (discussed further in Sec. \ref{sec:problems}), we finally chose to train the model using a subset of the dataset’s categories. We selected two representative types of textures from each group and built a training dataset containing fourteen classes, to ensure the model could learn a sufficiently diverse range of features, thus enhancing its capacity for generalized knowledge. The selected classes are shown in Table \ref{tab:training}.

\begin{table}\centering
  \caption{Selected classes of training set.}
  \label{tab:training}
  \begin{tabular}{cc}
    \toprule
    Group&Class\\
    \midrule
    G1&Rhomb Aluminum Mesh, Squared Aluminum Mesh \\
    G2&Granite Type Veneziano, Roof Tile \\
    G3&Aluminum Plate, Ceramic Plate \\
    G4&Bamboo, Laminated Wood \\
    G6&Carpet, Fiber \\
    G8&Cardboard, Wall Paper \\
    G9&Floor Cloth, Kashimir \\
  \bottomrule
\end{tabular}
\end{table}

First, a 3rd-order Butterworth bandpass filter was applied to extract frequency components ranging from 20 Hz to 1000 Hz. Subsequently, the Short Time Fourier Transform (STFT) was employed to construct spectrograms from the audio recordings, with a sampling rate of 44100 Hz. The STFT frame length was configured to 2048, accompanied by a hop length of 0.1 times the frame length. A Hann window was chosen as the windowing function.

To transform spectrograms back into sound waves to ensure that the vibrator can play them, we adopted Griffin-Lim algorithm \cite{griffin1984signal}. The number of samples per frame was set at 2048; the iteration time was set to 50; the hop length was set to 2048 / 10 (floor rounding). To reduce the time cost and provide a smooth interaction experience, we employed the algorithm on GPU using Torchaudio \cite{yang2021torchaudio}.

To train the model effectively, we segmented the spectrogram into fragments using a sliding window. This approach yielded several advantages, including dimensionality reduction and an augmentation in the training dataset's size. The dimensions of the sliding window were set at 48 units along the frequency axis and 320 units along the time axis, corresponding to a segmentation of 48 $\times$ 320 spectrograms. The choice of 48 on the frequency axis corresponded to 1000 Hz, signifying that only the segment of the spectrogram within this frequency range was retained. Concurrently, the 320 units along the time axis approximated a duration of 1.5 seconds for each spectrogram segment. This temporal length was deemed sufficient to capture the full information of the underlying audio signal.

The original dataset contains approximately 5 seconds sound recordings of different surfaces. Here, we used a sliding window of 48 $\times$ 320 corresponding to approximately 1.5 seconds, and moved it on transformed spectrograms, to ensure that we can obtain segmentations from different beginning time. The moving step length was set at five units (about 23 milliseconds), and produced a total of 24480 spectrogram segments.

\subsection{Model Training}
The size of the latent vector $z$ was set to 128. In our attempts, larger latent vector cannot improve the performance, but caused fluctuations and a longer convergence time. Network training hyperparameters were defined as follows: a batch size of 128, the learning rates of $G$ and $D_{s}$ were set to 0.0002, and the $E$, $C$ and $D_{z}$ were set to 0.001. The learning rates of $C$ and $D_{z}$ was decreased by a factor of 0.95 with the increasing of the epoch number. We opted for binary cross-entropy for the adversarial loss and cross-entropy for the auxiliary classification loss, while for the decoder loss, Mean Square Error (MSE) was chosen. The Adam optimizer was employed for efficient optimization. To mitigate overfitting, we utilized soft labels within the adversarial component. Specifically, a one-sided label smoothing was implemented, with a soft scale parameter set to 0.3. This signifies that random values between 0.7 and 1 were utilized instead of a fixed value of 1, and similarly, random values between 0 and 0.3 were employed instead of 0. The entire training regimen spanned 50 epochs. After the training process, only the generator part was used for the optimization system. The training process was conducted on an RTX A6000 GPU, and took 2.5 hours.

To expedite convergence, mitigate bias, and enhance stability, we applied min-max normalization to the training dataset, subsequently rescaling the range of values to fall within the range of [-1, 1]. This normalization process aligns the data range of both the input and output of the generator, facilitating the training process. Additionally, we performed the inverse transformation to revert the network's output range back to the original real data range, ensuring that the generated data remain consistent with the original data distribution.

\subsection{Initialization}
\label{sec:initialization}
An ideal initial point will help the optimizer find the target well. DSS has attempted initialization approaches, such as providing options or limiting the distance between initial point and target to control the difficulty of the task and avoid bad initialization \cite{Differential-Subspace-Search}.

There is a critical challenge in our optimization setup. While the auxiliary classifier enhances class distinguishability, it widens the gap between classes, potentially hindering the optimizer from traversing between clusters and restricting it to a confined space (details of the latent space distribution in Sec. \ref{sec:encoded_latent_space}). 

To address this challenge, we used a user option-based initialization method inspired by the tabu search \cite{GLOVER1986533}, a metaheuristic search technique that incorporates a ``prohibition" rule to facilitate solution discovery. Our straightforward initialization rule ensures a quick process.

A vibration of the training set is randomly chosen and transformed into the latent vector $z$ using the trained encoder, which serves as the initial value. Users compare the initial vibration with the target vibration, selecting a similarity rank from ``Good," ``So-so," or ``Bad." If the user chooses ``So-so" or ``Bad," another $z$ is selected until a ``Good" match is identified. The corresponding latent vector $z'$ is accepted as the initial value for the DSS optimizer.

To bring $z$ closer to the target area, the principle of our search strategy is: Choose near but different new $z$ for ``So-so", and far new $z$ for ``Bad". The average distance of vectors corresponding to the training set samples in the known latent space $dis_{avg}$ was calculated. Here, we define a parameter $step = \frac{1}{8} \times dis_{avg}$ as the unit of the moving distance during the exploration. Inspired by the tabu search, visited latent vectors are added to the tabu list. In this study, a specified radius $r$ is enforced for vectors within the tabu list (here $r$ is defined as the average distance $step$). Consequently, vectors within this range are prohibited, allowing for the exploration of larger regions. In this study, the length of the tabu list is set to five, meaning that the areas surrounding the five most recently visited latent vectors are prohibited from the further exploration. Following the search principle mentioned, when the user selects ``So-so", we choose a new $z$ within the range of (0.5 $\times$ $step$, 2 $\times$ $step$) from the current $z$. If the user chooses ``Bad", we select a new $z$ from distances beyond 2 $\times$ $step$ from the current $z$. 

This simple rule is easy to implement and allows for rapid execution. Through this method, we can efficiently select an initial latent vector that is closer to the target region, mitigating the risk of not reaching the target area owing to gaps in the distribution of classes.

\begin{figure}[h]
\centering
\subfigure[]{\includegraphics[width = 0.45\linewidth]{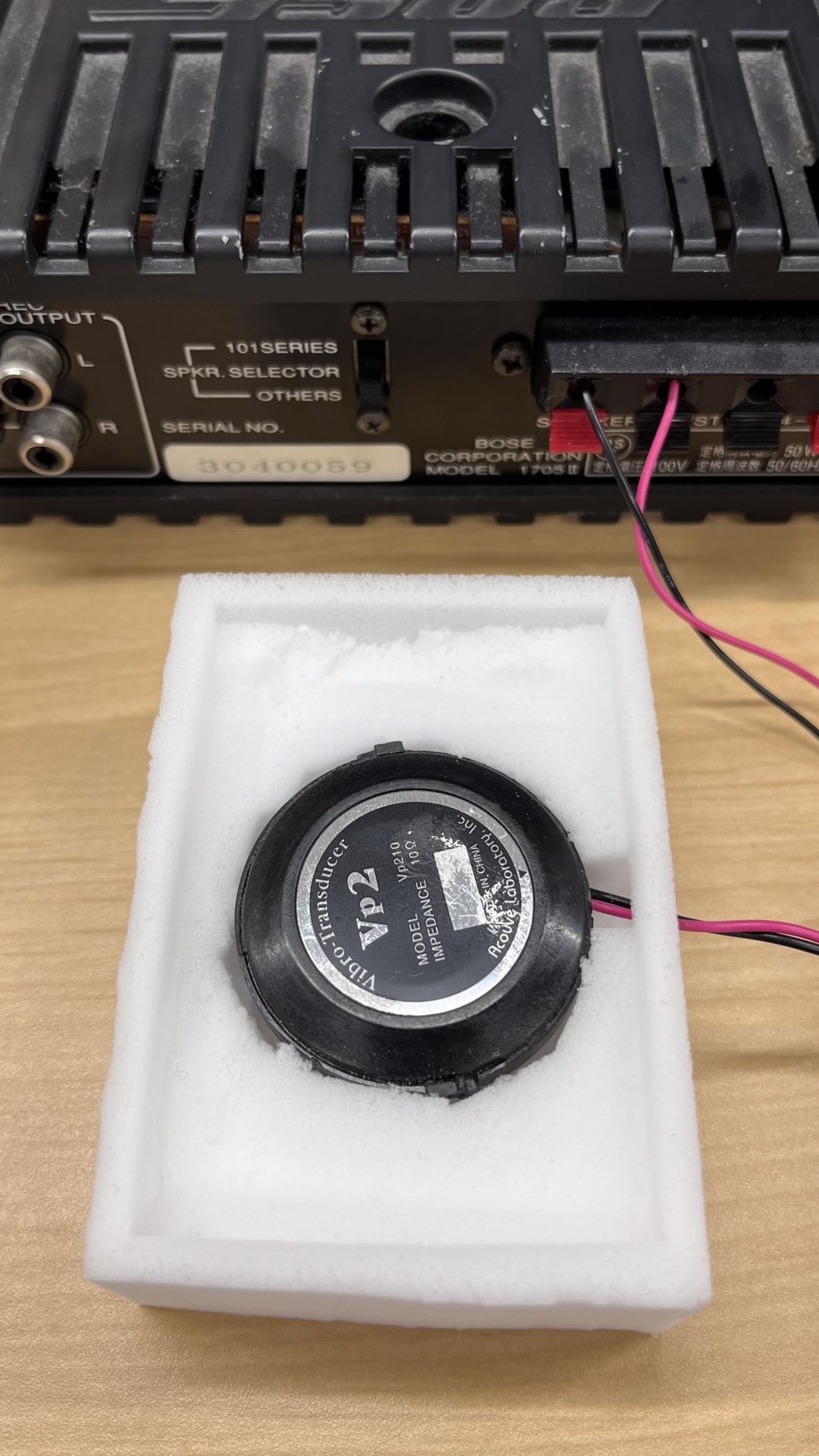}}
\subfigure[]{\includegraphics[width = 0.45\linewidth]{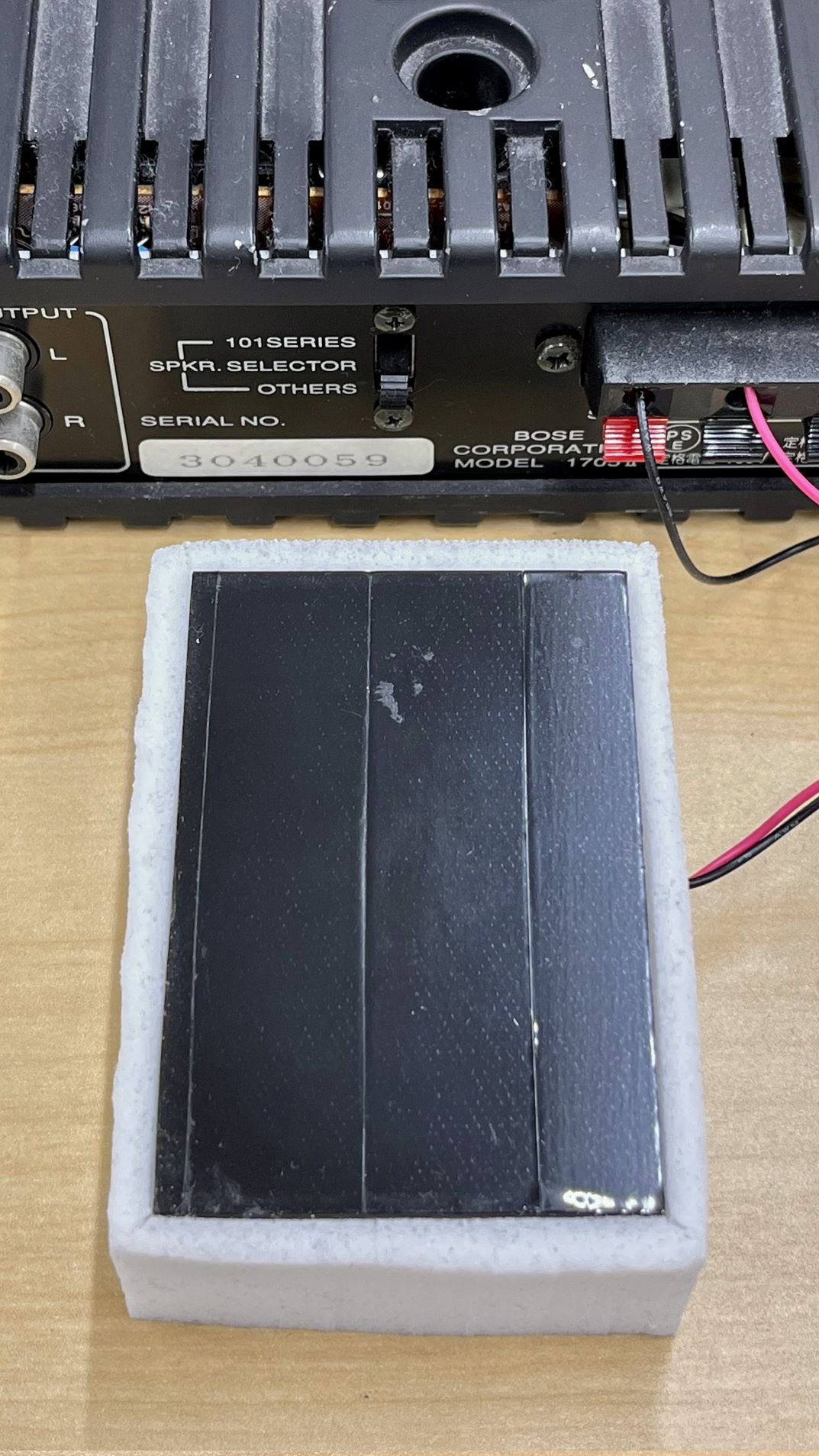}}
\caption{The vibrotactile presenting device in this research. The vibrator was placed in a foam block as shown in (a), and a piece of acrylic board was attached to the top of the vibrator for users to touch, as shown in (b).}
\label{fig:device}
\end{figure}

\subsection{User Interface and Presenting Device}
\begin{figure}[h]
  \centering
  \includegraphics[width=0.6\linewidth]{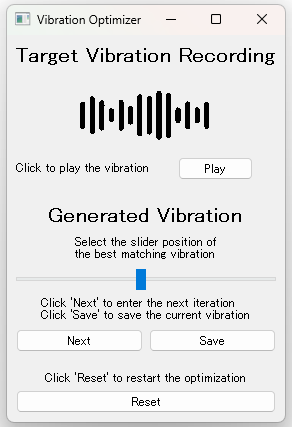}
  \caption{The user interface of the optimization.}
  \label{fig:UI}
\end{figure}

Fig. \ref{fig:device} and Fig. \ref{fig:UI} show the vibrotactile display used for evaluating the designed vibration and the user interface of the DSS. The device \cite{10198337} consists of a vibro-transducer (Acouve Vp210), a power amplifier (BOSE 1705II), foam, and an acrylic cover board. The vibrator was set in the foam block to ensure that the vibration can be transmitted effectively. An acrylic plate was attached to the vibrator, and a layer of high-friction electrical tape was applied to the upper surface of the plate. This adjustment enhanced the friction, ensuring participants' fingers maintained closer contact with the vibrating surface and enabling them to perceive the vibrations more effectively.

During the optimization process, users can click the ``Play" button above to play the actual vibration recording serving as the target. When users drag the slider below, vibrations generated from the feature vector corresponding to the slider's position will be played. Users can freely explore the high-dimensional feature space by moving the one-dimensional slider. When users locate the slider position that best matches the target vibration, they can click ``Next" to proceed to the next iteration. Users can also click ``Save" at any time to save the currently generated vibration.


\subsection{Stage I: Generation}
\begin{figure}[tb]
  \centering
  \includegraphics[width=\linewidth]{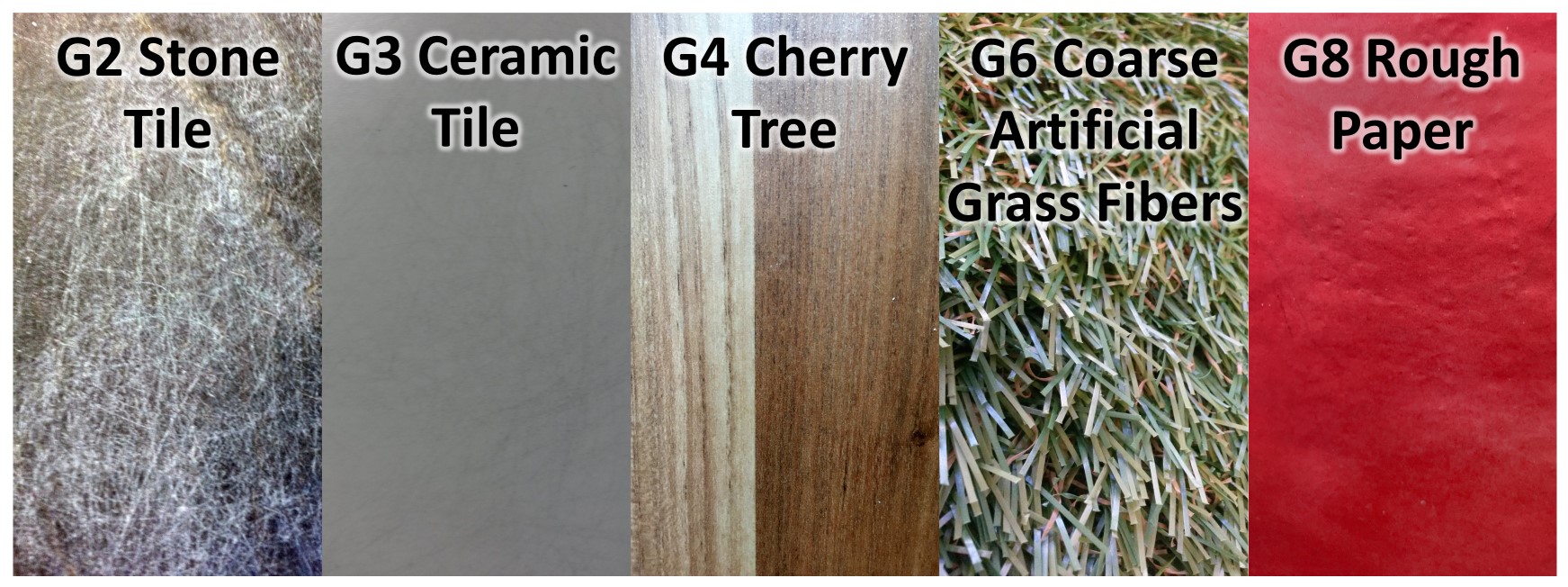}
  \caption{The figure shows selected 5 target classes in the LMT108 dataset to generate.}
  \label{fig:target}
\end{figure}

To assess the generative performance of our TexSenseGAN, we designed a two-stage experiment that includes a generation experiment and an evaluation experiment. To validate whether TexSenseGAN can generate new samples different from the training data, we selected classes different from the training set as generation targets. Although we followed previous research and used 7 groups in the training process, our further observation was that G1 and G9 exhibited excessively weak vibration intensities or significant similarity to other categories from the subjective feeling. Consequently, we also excluded G1 and G9. Finally, following the principles of stratified sampling, we randomly selected one class from each of the remaining groups for the task. The selected classes of the LMT108 dataset are shown in Fig. \ref{fig:target}.

Owing to the free movement of the collection device on surfaces during the vibration acquisition process \cite{strese2016multimodal}, vibration is not uniformly distributed on the time axis. Therefore, not all samples can effectively represent the characteristics of a particular class. To provide subjects with more reliable target vibrations, facilitating better execution of the generation task, we manually selected three relatively uniform and representative vibrations for each class. This resulted in a total of fifteen target vibrations, and subjects were tasked with generating similar samples to these vibrations.

In comparison to some slider-based human-in-the-loop visual optimization tasks \cite{Sequential-Line-Search, Differential-Subspace-Search}, the haptic perception is not as sensitive as vision. To help subjects better understand how to perceive vibration characteristics and the differences between vibrations of different categories, we conducted some training tasks before the main experiment. In this process, we informed subjects about the system's operation, showed them target vibrations from different classes to help them understand the distinctions, presented target vibrations from the same class to illustrate commonalities, and allowed subjects to perform actual operations to familiarize themselves with the experimental procedure and comprehend our optimization goals.

After subjects familiarized themselves with the system and characteristics of vibrations, we proceeded to the generation experiment process. Subjects were required to generate a sample corresponding to each target vibration. The first step was the initialization of the optimization process. In this step, subjects needed to use the options ``Bad", ``So-so", and ``Good" to select a suitable initial value for the optimizer. During this process, subjects were only required to make rough comparisons; thus, they were instructed not to focus on detailed perceptions, but to perform quick and multiple selections, with the aim of covering an exploration range through multiple attempts. It is important to note that the three options here do not represent the actual quality of the samples. For instance, “Good” does not mean that the sample is sufficiently close to the target sample, but rather that it possesses certain characteristics similar to the target (e.g., similar frequency components), making it a suitable initial point for optimization. After entering the optimization process, subjects were encouraged to make numerous selections rather than sticking to a single iteration's slider position selection. 

\begin{figure}[ht]
  \centering
  \includegraphics[width=\linewidth]{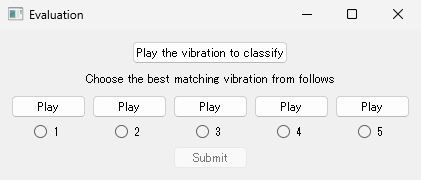}
  \caption{The user interface of the Stage II. Subjects can click the upper button to play the vibration to classify. Then subjects should select a best matching one from the below real vibration samples.}
  \label{fig:EvaluationUI}
\end{figure}

However, in this phase, subjects were asked to carefully perceive various aspects of vibrations, such as frequency and amplitude. We introduced several effective ways of perceiving signal changes and adjustment strategies to help participants better understand the generation process and complete the task more easily. Participants are encouraged to adjust one parameter at a time, such as frequency, before proceeding to other aspects. This approach enhances the clarity of changes during the tuning process. To ensure optimal results, participants should handle the sliders carefully, as excessive adjustments may produce distorted vibration signals. Such distortions often indicate that the adjustments have pushed beyond the model's acceptable latent space, potentially leading to generation failures. Therefore, avoiding these results is crucial. Moreover, to distinguish between different slider positions, this optimization process might be more time-consuming. When consecutive iterations did not produce vibrations significantly different along the slider direction, subjects could adopt several strategies:

\begin{itemize}
  \item [1)] 
  Continue selecting similar slider positions as much as possible.     
  \item [2)]
  Randomly choose slider positions to help the optimizer move away from the current situation.  
  \item [3)]
  If the subject felt dissatisfied with the current vibration, restart the optimization process from the initialized point. 
\end{itemize}

Following strategy 1 or strategy 2 for several consecutive iterations usually allowed the optimizer to move away from the current state. During the experiment process, subjects were asked to freely choose strategies based on their subjective judgment. However, we encourage participants to primarily opt for strategies 1 or 2 and minimize the use of the reset operation during the optimization process. If participants feel fatigued, prefer not to continue with the iterative adjustments of strategies 1 or 2, or if the generated samples become distorted, they may then choose strategy 3. Since carefully perceiving subtle differences in vibration can be mentally exhausting, and prolonged exposure may cause finger numbness, potentially affecting sensitivity, participants are allowed to take breaks at any point during the experiment to maintain accuracy in their perceptions. In our experimental observations, reset operation was rarely used; the subject who used the reset operation the most did so only three times throughout the entire task, and rest subjects almost never used it. 

During the selection of initial points and the optimization of vibrations, participants were also asked to rate the subjective similarity between the sample and the target at three different time points: the initial random sample, the sample after initialization, and the optimized sample. Ratings were given on a scale of 0 to 10 (0 = completely different, 10 = highly similar). Unlike the options in the initialization mentioned earlier, participants here were required to directly assess the degree of similarity between the current sample and the target. The entire process of the generation experiment lasted approximately 2 hours, including rest breaks. Each subject generated 15 samples corresponding to the target vibrations. From initialization to completion, the generation of each sample took 2-8 minutes.

To cover multiple generated results in the subsequent pairing task of the evaluation experiment without overwhelming the subjects, we limited the number of generated samples. In the first stage, we invited three subjects (three males, aged 21-26), who generated 45 samples (nine for each class). These three subjects will not be involved in the evaluation phase to eliminate the impact of memory.

\begin{figure}[tb]
  \centering
  \includegraphics[width=\linewidth]{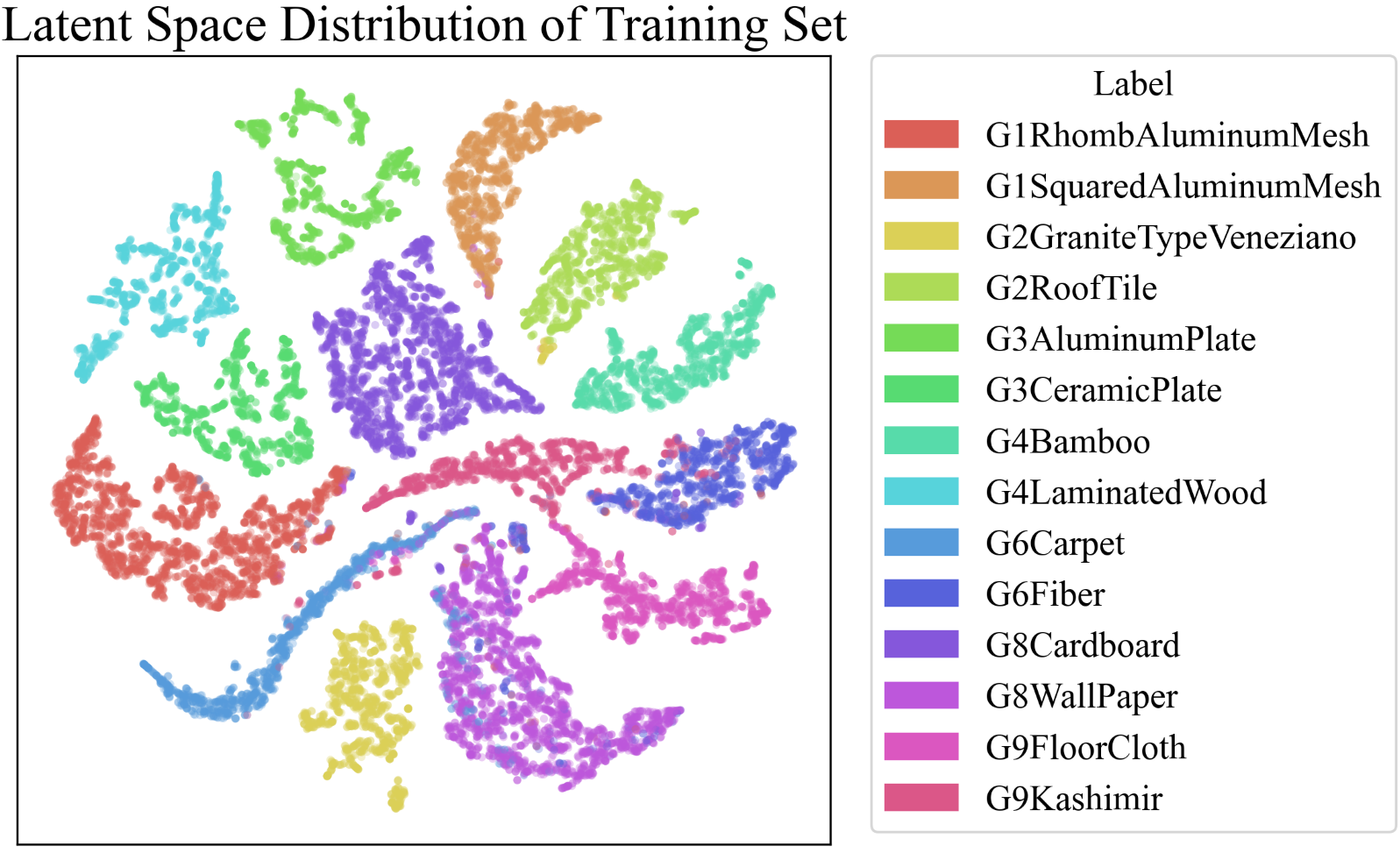}
  \caption{The distribution visualization of the latent space.}
  \label{fig:latent}
\end{figure}

\subsection{Stage II: Evaluation}
We designed a classification task to validate the similarity between the vibrations generated and the real vibrations of the surface. The user interface for this part is shown in Fig. \ref{fig:EvaluationUI}. The classification task comprised three parts:

\begin{figure*}[h]
  \centering
  \includegraphics[width=\linewidth]{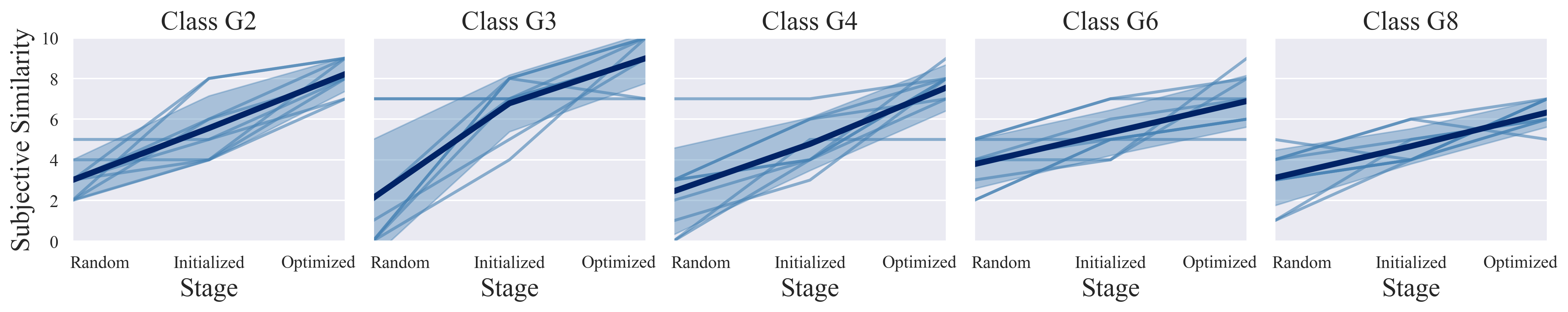}
  \caption{The figure shows the subjective similarity of each vibration class in different optimization stages.}
  \label{fig:SubjectiveSimilarity}
\end{figure*}

\begin{figure*}[h]
  \centering
  \includegraphics[width=\linewidth]{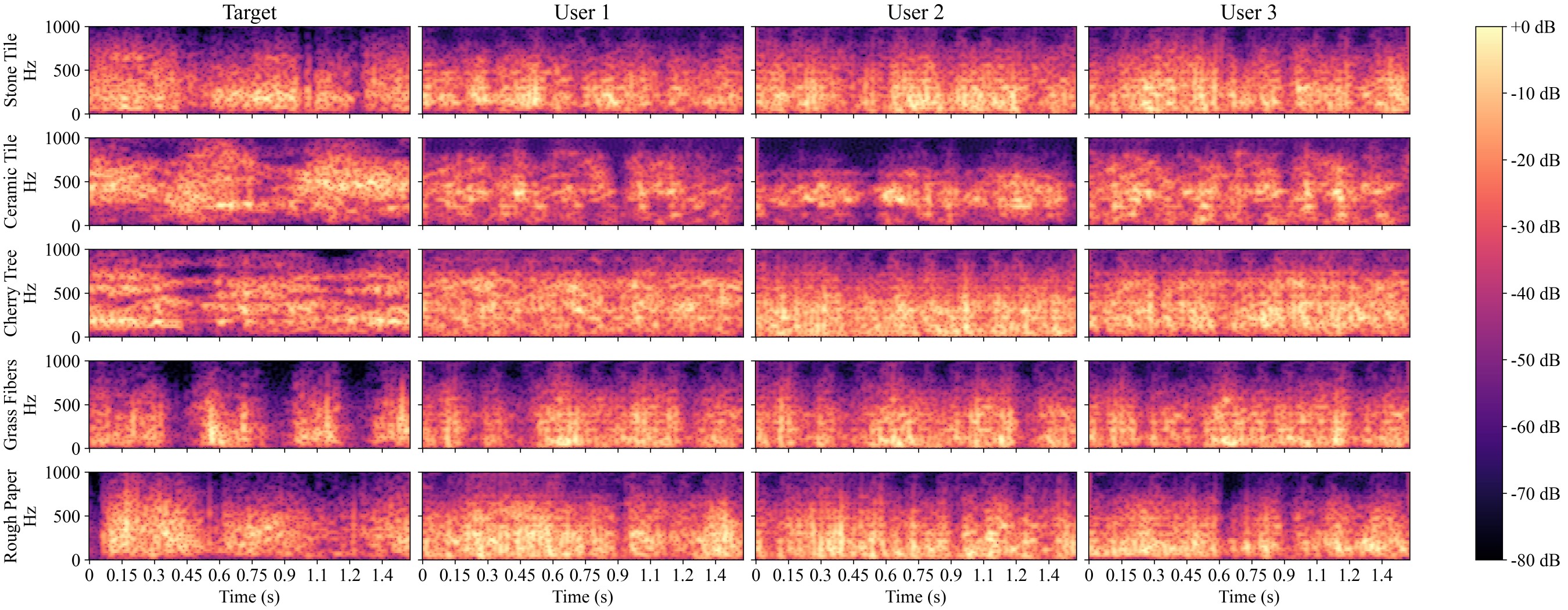}
  \caption{The figures show one target spectrogram for each class, along with the corresponding spectrograms generated by three users. The leftmost column represents the real target vibrations used as references, while the three columns on the right each represent one of the vibrations generated by individual users. Each row corresponds to a different vibration class.}
  \label{fig:GenerationResults}
\end{figure*}

\begin{figure*}[h]
  \centering
  \includegraphics[width=\linewidth]{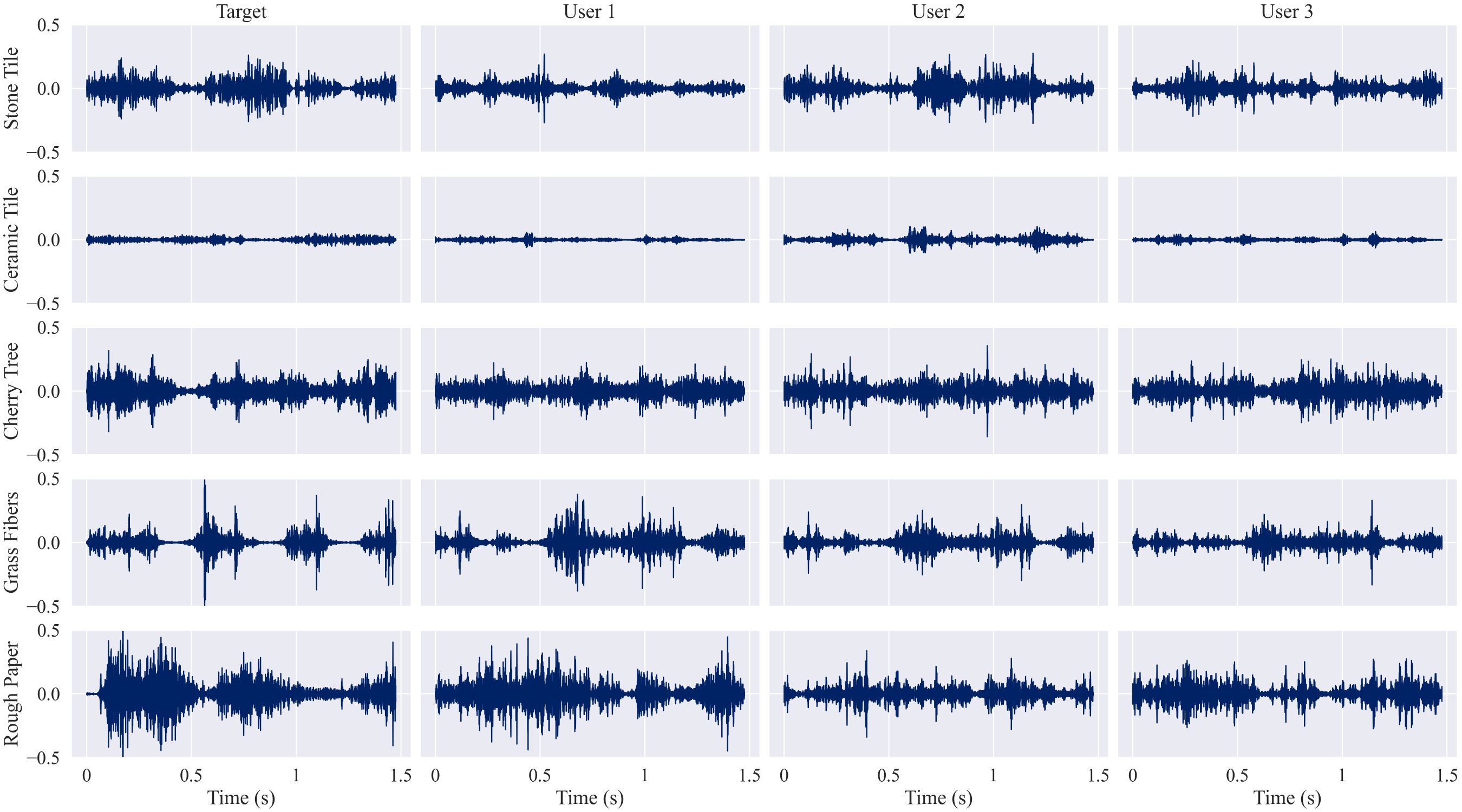}
  \caption{The figures show the waveforms of the vibrations corresponding to the spectrograms in Fig. \ref{fig:GenerationResults}}
  \label{fig:GenerationResults_wave}
\end{figure*}

\begin{figure}[h]
  \centering
  \includegraphics[width=\linewidth]{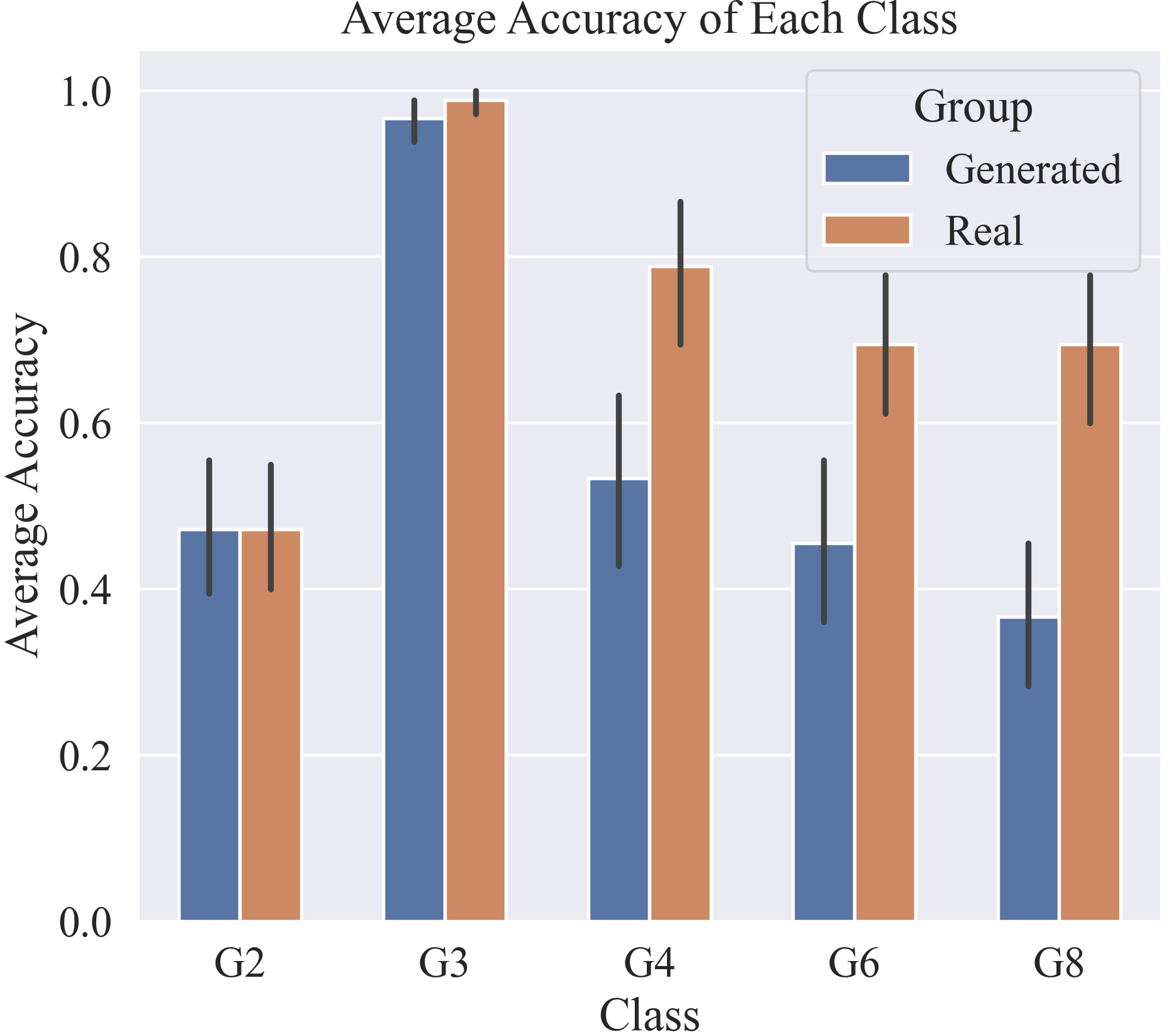}
  \caption{The average accuracy was calculated in each class.}
  \label{fig:ClassAccuracy}
\end{figure}

1. Select the correct class for the real vibration (the control group): In this part, all 15 real vibrations were presented in a shuffled order to the subject. For each trial, 1 real vibration was selected for classification, and subjects could click a button to play this vibration. Furthermore, 5 real vibrations from each class were presented. These were randomly selected from each class, excluding the vibration to be classified shown above. The indices (1, 2, 3, 4, 5) corresponding to the vibration classes (G2, G3, G4, G6, G8). Subjects could play these vibrations and were required to choose the class that most closely resembled the one being classified. This process was repeated three times, resulting in a total of 45 classifications. This part aimed to validate the study's premise that people can perceive differences between vibration stimuli, understand their characteristics, and accurately classify them. The results from this control group were used to assess the accuracy of the generated vibrations. Subjects were asked to describe the features of the five types of vibrations and perform subsequent tasks based on these descriptions.

2. Select the correct class for the generated vibration. The procedure for this part was similar to the first, but focused on classifying generated vibrations. The forty-five generated vibrations were shuffled and presented to the subjects for classification. The vibrations below were randomly selected from the real vibrations of each class. Subjects were required to select the class that most closely resembled the vibration generated from these five options. This process was only conducted once to maintain an equal task number with the control group. We asked subjects to describe the vibration features and compare with the control group.

3. Select the corresponding target vibration for the vibration generated. The third part aimed to validate whether subjects can identify differences between samples in the same class. The forty-five shuffled generated vibrations were shown to the subjects, same as in the second part above. However, instead of five options, subjects were presented with three options, all belonging to the same class as the displayed generated vibration. This is because, for each class, there were three real vibrations serving as targets. Therefore, subjects needed to select the correct target vibration from below three options corresponding to the given generated vibration. The total number of classification tasks in this part was also 45, determined by the number of vibrations generated.

Twenty subjects (eleven males, nine females, aged 22-27) took part in this stage. The entire process of the evaluation experiment lasted about 1.5 hours, including the tutorial and rest breaks.

Approval of all ethical and experimental procedures and protocols was granted by the Ethics Committee of The University of Tokyo.

\section{Results}
In this section, we will first present the performance of the trained neural network model, and then present the results of the two-stage user experiments.

\subsection{Encoded Latent Space}
\label{sec:encoded_latent_space}
The latent space is shown in Fig. \ref{fig:latent} after the 50-epoch training process. With the benefit of the auxiliary classification, the latent space can be distinguished according to the class.

\subsection{Problems in Other Alternatives}
\label{sec:problems}
In addition to the network model used in TexSenseGAN, we implemented various generators or decoders for comparison, including basic GAN, autoencoder; Variational Autoencoder (VAE) to learn the distribution; AAE to align the distribution, CGAN, and ACGAN to incorporate conditional information. However, these models faced challenges operating within our optimization framework, making direct comparisons unfeasible here. 

Initially, we experimented with models based on the autoencoder architecture, including autoencoders, VAEs, and AAEs. While these models were relatively easy to train, they suffered from severe over-smoothing issues. Considering our need for feature vectors suitable for optimization, we targeted a relatively small feature space dimension, such as the 128 dimensions used in this study. However, this dimensionality proved inadequate for the original data, resulting in significant information loss when encoded into this latent space. Consequently, the generated spectrograms lacked detailed information, leading to noticeable differences between the reconstructed vibrations and the originals.

Moreover, we incorporated GANs into our framework. Although GANs could generate spectrograms with more detail compared to autoencoders, they were prone to mode collapse, producing similar results for different inputs. Thus, GANs were unsuitable for our optimization approach based on input vectors. Additionally, GANs generate results from random noise, resulting in a mixed latent space. Consequently, the generated vibrations underwent significant changes when users moved the slider continuously. Contrary to the original DSS research, which focused on visual stimuli, our work emphasized on tactile tasks, which were less sensitive than visual stimuli. Consequently, these abrupt changes made it difficult for users to touch and compare vibrations, complicating the task of selecting an appropriate slider position.

\begin{figure*}[ht]
\centering
\subfigure[]{
    \includegraphics[height=5.7cm]{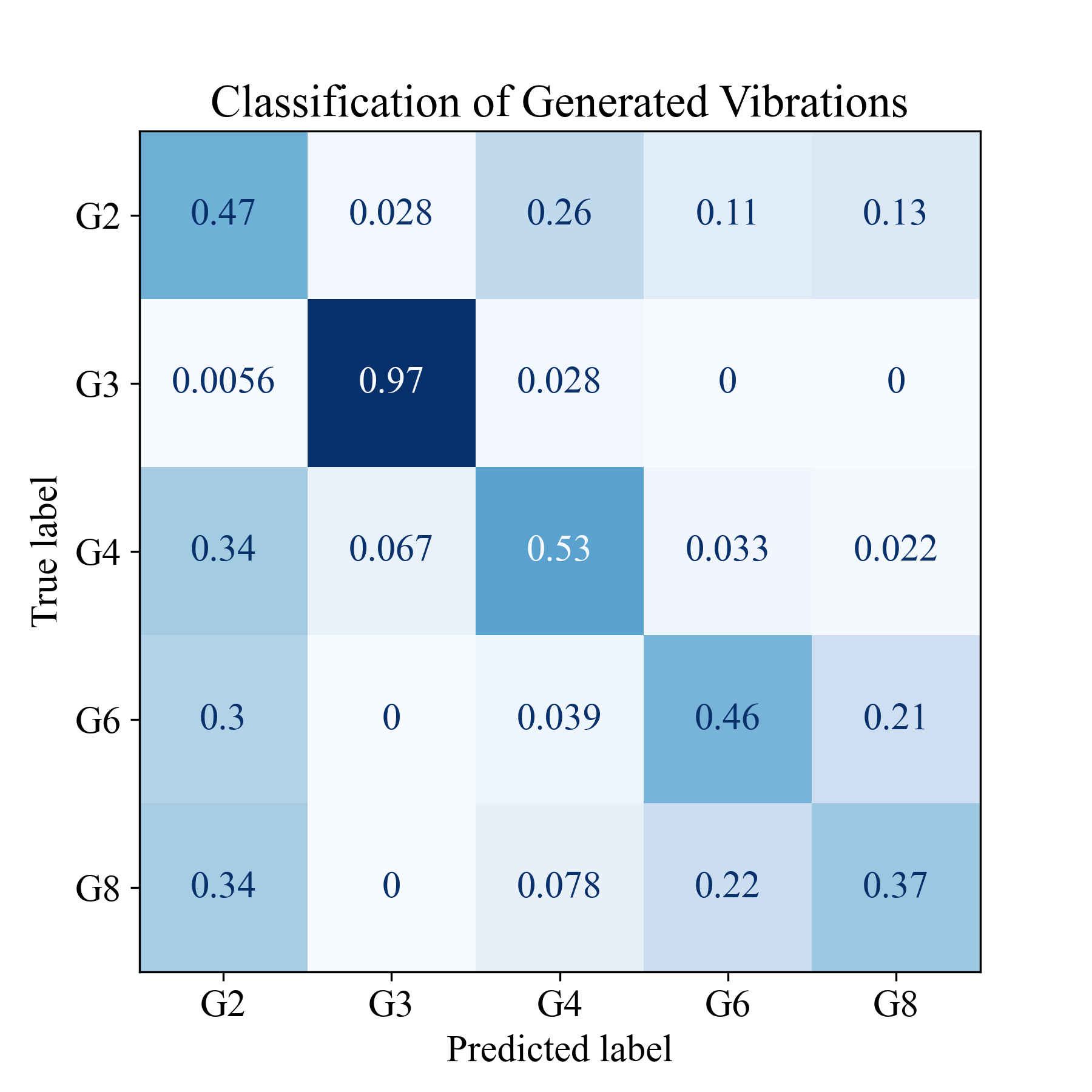}
    \label{fig:ClassificationG}
}
\subfigure[]{
    \includegraphics[height=5.7cm]{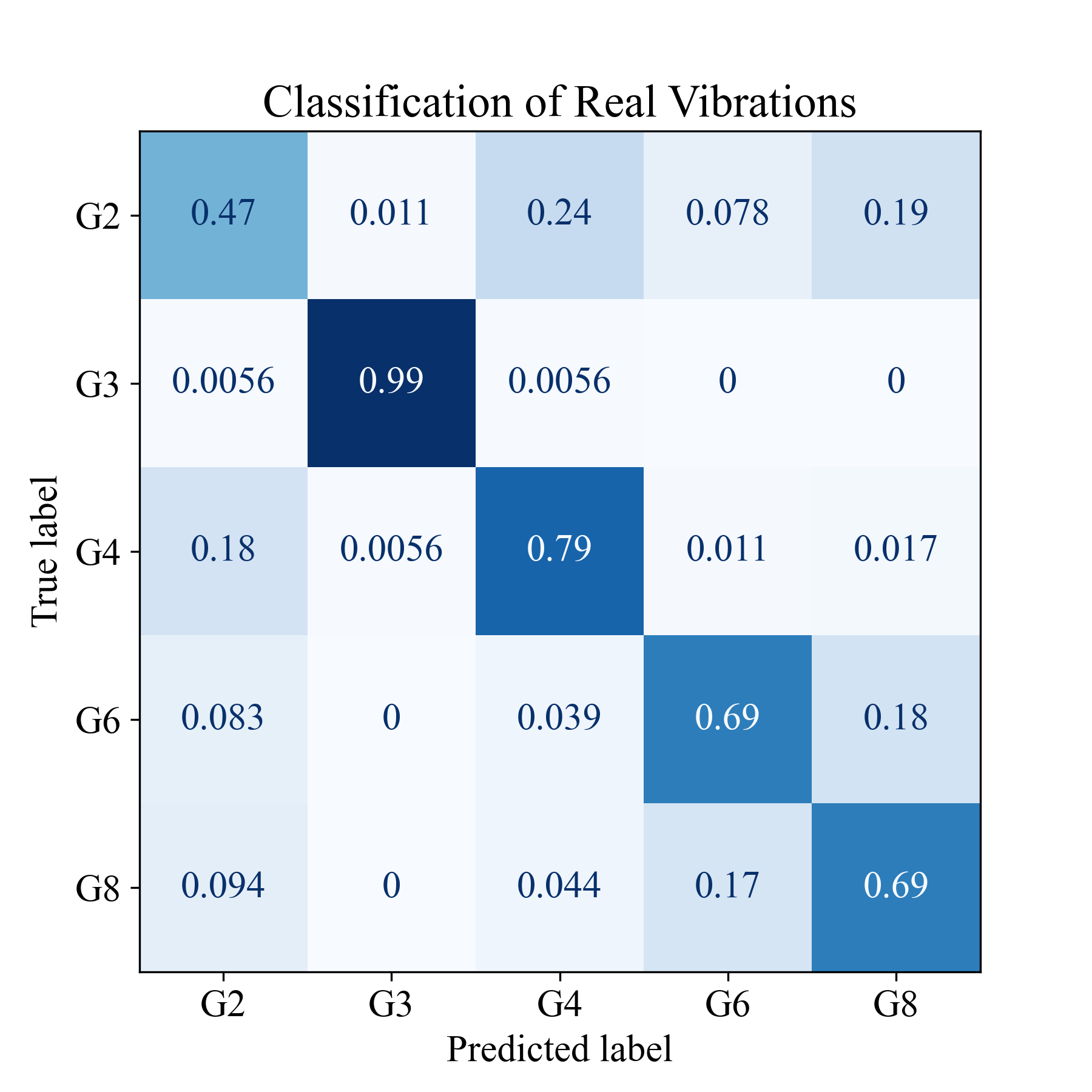}
    \label{fig:ClassificationR}
}
\subfigure[]{
    \includegraphics[height=5.7cm]{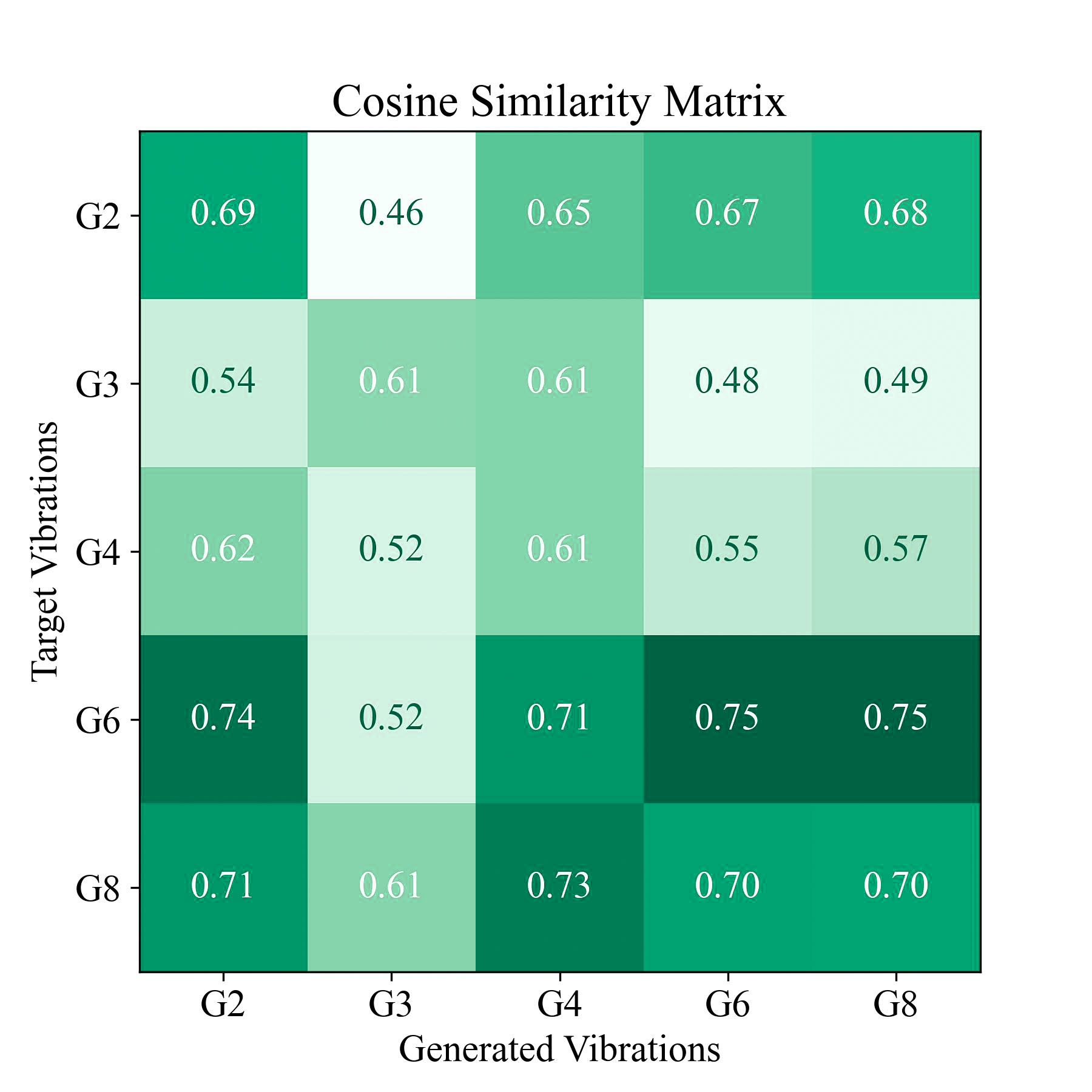}
    \label{fig:ObjectiveSimilarity}
}
\caption{(a) The confusion matrix of the classification task of generated vibrations, (b) The confusion matrix of the classification task of real vibrations, (c) \hl{The average cosine similarity matrix between generated samples and target samples.}}
\end{figure*}

We compared generative models, CGAN and ACGAN, with conditional information support. Within our task, CGAN only requires conditional information as input, thus avoiding the challenges of multi-class auxiliary classification in ACGAN, allowing it to handle more complex datasets. However, our tests demonstrated that this approach of using conditional information alongside latent vectors as input to control generation, was less effective than ACGAN’s direct use of an auxiliary classifier to shape the latent space distribution. Under our training framework, CGAN tended to produce extremely sparse output results or exhibited a tendency not to converge. ACGAN performed relatively better among the control group models we experimented with; however, it faced challenges when applied in the optimization process, such as DSS. ACGAN can differentiate features between different classes and control them, but the uneven input vector format of label + latent vector is not conducive to the optimization process of DSS. Despite these challenges with ACGAN, we incorporated a similar auxiliary classification approach in TexSenseGAN, leveraging its strong capability for latent space control.

These problems make it difficult to identify models with similar performance for comparison with our TexSenseGAN. Therefore, we excluded the performance results or user studies of the control group.

\subsection{Subjective Similarity and Generation Results}
Fig. \ref{fig:SubjectiveSimilarity} illustrates changes in subjective similarity to the target sample across three stages of the generation process: ``Random" represents the initial random latent vector, ``Initialized" denotes samples that have been initialized with certain target characteristics, and ``Optimized" indicates the final optimized result. Light thin lines show changes in each individual trial (9 trials per class, with 3 participants generating samples for 3 target samples). The dark thick line represents the average subjective similarity for each class, with error bands indicated. Overall, the optimization process demonstrates effectiveness, supported by the benefits of the initial value selection method. Nevertheless, certain cases reveal instances of ineffective or even inverse optimization, with variations in effectiveness across different classes. Also, some generated spectrogram samples of Stage I are shown in Fig. \ref{fig:GenerationResults}, and the corresponding waveforms transformed from the spectrograms are shown in Fig. \ref{fig:GenerationResults_wave}, to show the generation performance.

In the figure, the leftmost column shows five categories of target vibrations, while the three columns on the right show the vibrations generated by three users for these targets. We selected one vibration from each category for demonstration. The spectrograms reveal that our TexSenseGAN successfully produces vibration patterns similar to the target for certain categories. While some results are less ideal, the tendency to generate more uniform vibration patterns shows the consistency of the system. This consistency is a strength, although it can make replicating vibrations with significant temporal fluctuations, such as Grass Fiber, more challenging. Moreover, the GAN structure significantly aids in generation; however, some detail loss owing to the autoencoder-like structure results in smoother outputs compared to real vibrations. Nonetheless, the system effectively replicates the general vibration characteristics to a certain extent, even if the intensity is slightly reduced compared to the original samples, such as Rough Paper.

However, spectrograms alone cannot fully capture the subjective experiences of the subjects. Therefore, we mainly focused on analyzing the results of Stage II, the classification task. 

\subsection{Inter-class Classification}
From Fig. \ref{fig:ClassAccuracy}, we can observe an overview of the result of the inter-class classification task. In the five-category task, subjects achieved an accuracy of 55.9\% when judging the generated vibrations (experimental group, experiment 2), compared to approximately 72.8\% for the real vibrations (control group, experiment 1). The control group results indicate that subjects can distinguish between different surface vibration stimuli. Fig. \ref{fig:ClassAccuracy} shows that subjects in the generation group and the control group exhibited similar accuracy trends for these 5 types of different vibrations. This preliminary evidence suggests that our TexSenseGAN can generate distinct vibration patterns that subjects can differentiate. 

Fig. \ref{fig:ClassificationG} shows the confusion matrix of the classification task of generated vibrations and Fig. \ref{fig:ClassificationR} shows the control group, the classification of real vibrations. The confusion matrices provide a clearer view of how subjects distinguished between each type of vibration. We can observe that both 2 confusion matrices show a diagonal trend. 

The best generation was achieved with G3 vibrations, with a classification accuracy of 97\% in the experimental group. The spectrogram shows that G3 has a higher proportion of high-frequency components compared to other categories. Subjects generally described G3 as a uniform and subtle high-frequency vibration. This distinct vibration pattern made it easier for subjects in Stage I to optimize the generated vibrations and for subjects in Stage II to distinguish this type of vibration.

Observing the confusion matrices, we find that their trends are similar but there are some differences in accuracy appearing in categories G4 (26\%), G6 (23\%), and G8 (32\%). In the experimental group, subjects classified the generated G4 samples with an accuracy of 53\%, G6 samples with an accuracy of 46\% and G8 samples with an accuracy of 37\%, while in the control group, the accuracy for G4, G6, and G8 was 79\%, 69\%, and 69\%, respectively. This is because more G4, G6, and G8 samples were classified as G2. According to subject feedback, the characteristics of G2 vibrations are less distinct compared to other categories. Subjects often described G2 as a uniform and smooth mid-low frequency vibration or found it difficult to describe any prominent patterns. It can be observed that even in the control group, distinguishing the real G2 vibrations was challenging, with an accuracy of 47\%, same as the generated group.

\subsection{\hl{Subjective Classification and Mathematical Similarity}}
\hl{To assess the effectiveness of subjective judgment as the primary guide for adjustment, we computed the objective similarity between generated and target samples as a supplementary measure. Since our primary concern is whether the frequency components of the samples are similar, and given that the generated and target samples are not entirely identical (e.g., differences in starting time and phase), we applied a Fourier transform to all samples and extracted only their magnitude spectra. For each generated sample, we computed cosine similarity with the magnitude spectra of all target samples. The resulting pairwise similarities were averaged to form the matrix shown in Fig. \ref{fig:ObjectiveSimilarity}.

Comparing Fig. \ref{fig:ObjectiveSimilarity} with Fig. \ref{fig:ClassificationG} and Fig. \ref{fig:ClassificationR}, the generated samples perform noticeably worse on objective metrics than on subjective classification results. However, we also observe that for some easily confused categories, the objective metrics exhibit trends similar to those seen in subjective classification, indicating partial agreement between mathematical and perceptual measures. In other words, the trend of similarity provides a common ground between subjective and mathematical indicators, reinforcing each other. This phenomenon suggests that within the framework of this study, subjective perception is more sensitive than mathematical metrics, allowing us to obtain more meaningful and precise results.}

\begin{figure*}[h]
  \centering
  \includegraphics[width=\linewidth]{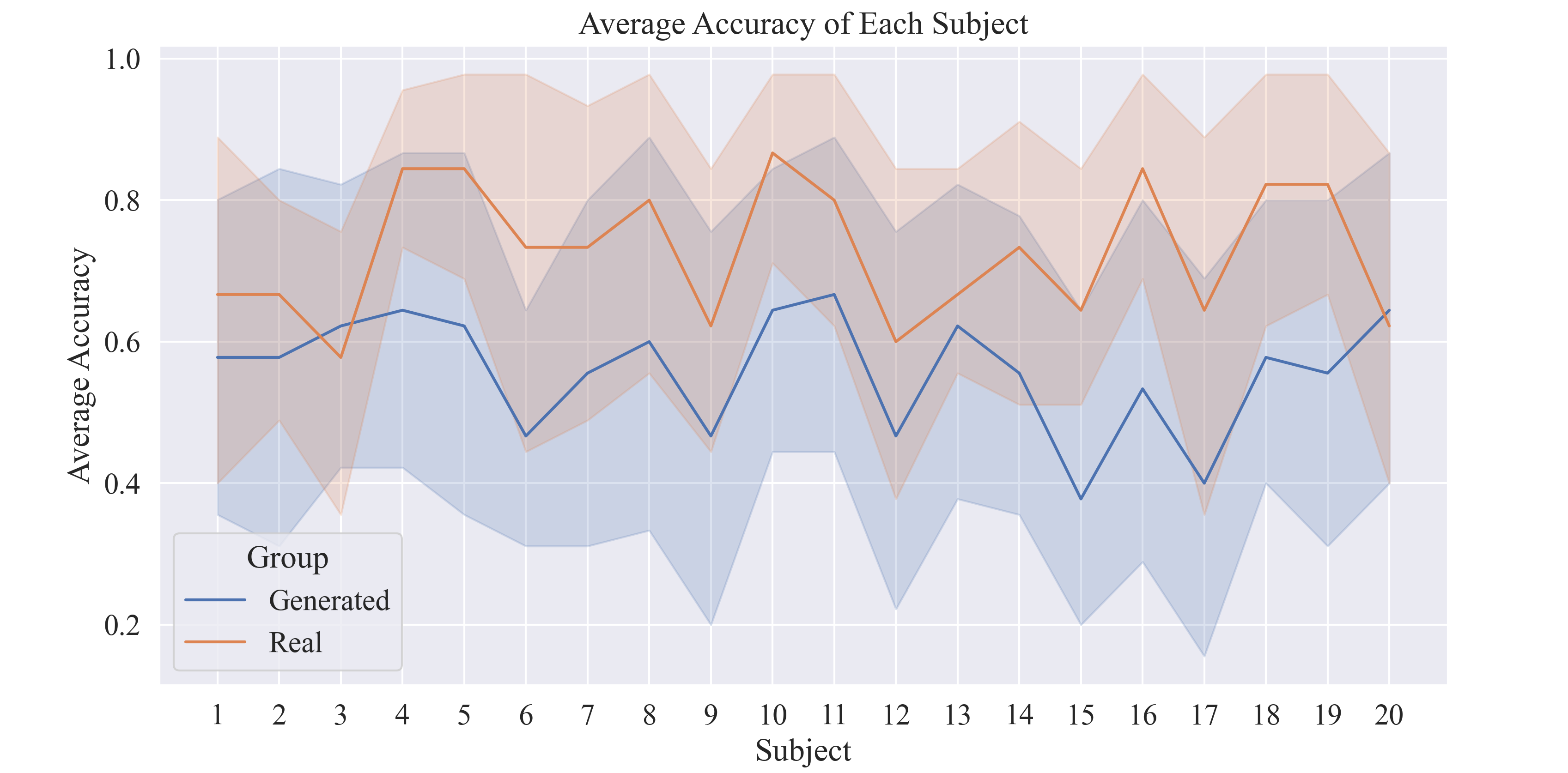}
  \caption{The average accuracy of each subject.}
  \label{fig:SubjectAccuracy}
\end{figure*}

\subsection{Regression Analysis}
We can draw some conclusions based on each subject's performance in both the generated and control group, as shown in Fig. \ref{fig:SubjectAccuracy}. We calculated the average accuracy of each subject for each class, obtained 100 sets of accuracy. The figure roughly shows an interesting trend: subjects who were good at distinguishing real vibrations in the control group also tended to have higher classification accuracy in the generated group. To further illustrate this phenomenon, we constructed a multivariate regression model following the process below to verify the relationship between the accuracy in the control group and the accuracy in the generated group.

We constructed a linear regression model with interaction terms, using control group accuracy, subject ID, and vibration class as independent variables, and the generated group accuracy as the response. Before the fit process, we removed outliers that exceeded two scaled median absolute deviations from the median, obtained 93 sets of accuracy. The regressed model can be expressed as: 

\begin{equation}
\begin{aligned}
acc_G = 0.0349 * c - 0.0017 * i + 0.9490 * acc_R + \\ 0.0011 * c * i - 0.1983 * c * acc_R - 0.0062 * i * acc_R.
\end{aligned}
\end{equation}

In this equation, $acc_G$ represents the generated group accuracy, $c$ represents the class ID, $i$ represents the subject ID, and $acc_R$ represents the control group accuracy. The $p$-values of each term are shown in Table \ref{tab:model}. The analysis reveals that the accuracy of the control group, $acc_R$, has the most significant impact, along with the interaction term between class $c$ and $acc_R$, both of which are statistically significant. This suggests a meaningful relationship between $acc_G$ and $acc_R$, indicating that our model successfully reconstructs recognizable features of real target samples. The significance of the interaction term $c * acc_R$ implies that the model's generation capability varies across different classes. Additionally, terms involving the subject ID $i$ are insignificant, suggesting that the observed patterns are consistent across subjects.

\begin{table}\centering
  \caption{Coefficients and $p$-values of each factor in the model.}
  \label{tab:model}
  \begin{tabular}{ccc}
    \toprule
    &Coefficient&$p$-value\\
    \midrule
    $c$     &0.0349  &0.5236 \\
    $i$     &-0.0017 &0.8963 \\
    $acc_R$ &0.9490 &8.4611$\times 10^{-5}$ \\
    $c * i$ &0.0011 &0.6998 \\
    $c * acc_R$ &-0.1983 &0.0053 \\
    $i * acc_R$ &-0.0062 &0.7012 \\
  \bottomrule
\end{tabular}
\end{table}

We performed a stepwise regression to select some variables. We can obtain a model like:

\begin{equation}
acc_G = 0.10245 * c + 1.0914 * acc_R - 0.2657 * c * acc_R.
\end{equation}

This model maintains class $c$ ($p = 0.0699$), control group accuracy $acc_R$ ($p = 1.1785 \times 10^{-8}$), and their interaction term $c * acc_R$ ($p = 0.0012$). The $acc_R$ is the most dominant term. The model shows the difference in generation ability between classes, even larger than the previous model. This is reasonable according to our experiment results. The model $R^2 = 0.423$, reveals that there is a correlation; $F = 21.7$ with $p = 1.19 \times 10^{-10}$, indicating that the model is credited.

\begin{figure}[h]
  \centering
  \includegraphics[width=\linewidth]{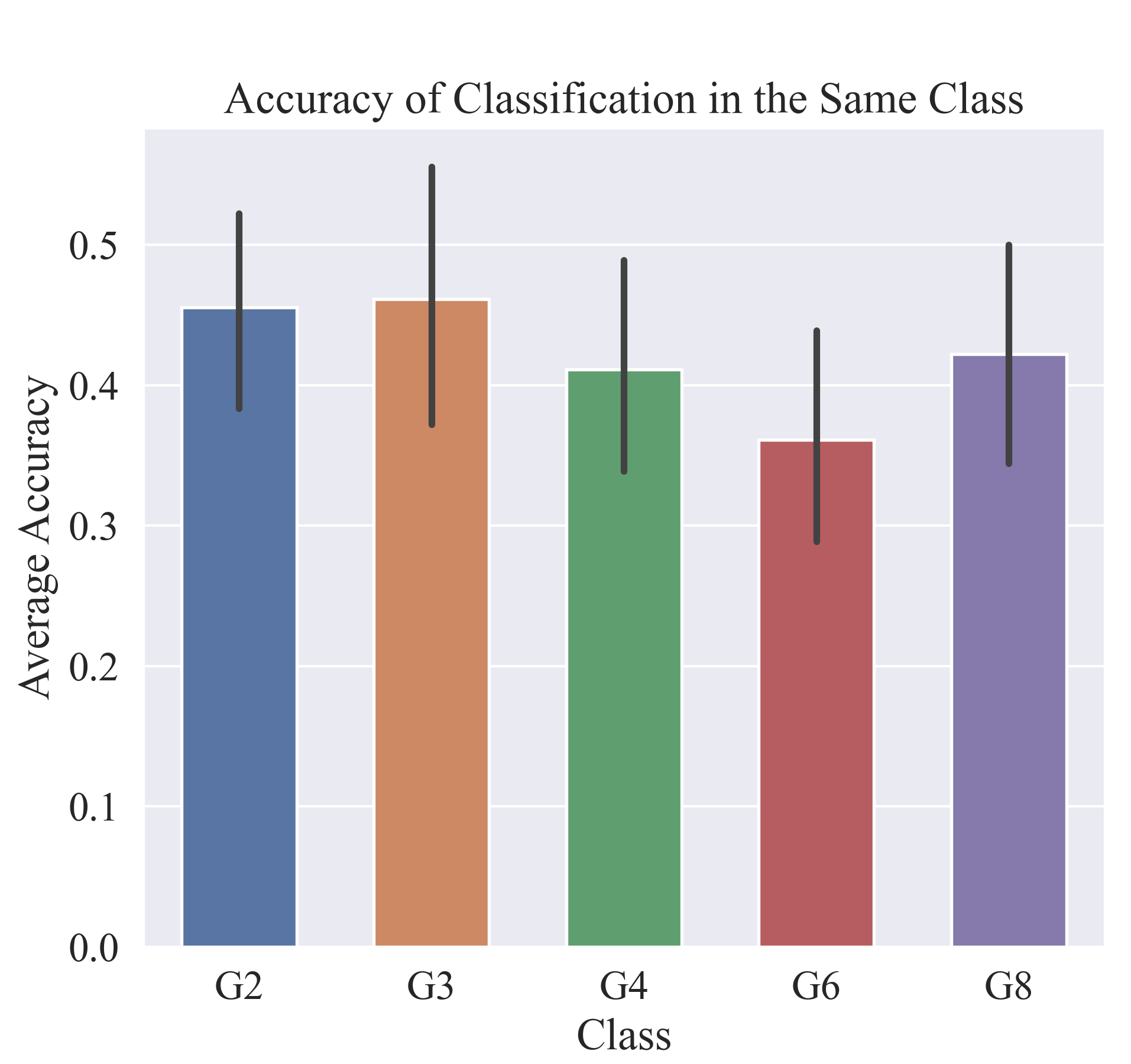}
  \caption{The in-class accuracy of each class. The accuracy of each class is just over 33\%, which means that although there are differences between the samples, they are difficult to distinguish.}
  \label{fig:inClassAccuracy}
\end{figure}

\subsection{In-class Classification}
Furthermore, we can obtain some information from the classification in a class. The overall 3-class classification accuracy is 42.2\%. This in-class accuracy of each class is shown in Fig. \ref{fig:inClassAccuracy}. The accuracy of each class is 45.6\% (G2), 46.1\% (G3), 41.1\% (G4), 36.1\% (G6) and 42.2\% (G8). Although the accuracy for each class is higher than the random classification for three classes (33.3\%), indicating that there are indeed some differences between samples generated with different targets, the classification accuracy is not sufficiently high to prove that users can distinguish between samples belonging to the same class. This can be caused by the performance limitation of the model on the reconstruction of detailed information or the limited haptic resolution. However, in another aspect, this result suggests that the model captures and replicates some essential and general characteristics that define each class. Despite challenges in distinguishing within a class, participants could distinguish between different classes. Therefore, the system reliably encodes and reproduces the defining features of each type of vibration.

\section{Discussion}
From the results, we observed several key phenomena that merit further discussion and analysis.

First, we can further discuss based on subjects' feedback and the patterns of these types of vibrations. The main frequency component of G4 vibrations is similar to G3 while the amplitude is stronger, similar to G2. Although the accuracy of real G4 achieves a high level, various subjects said G4 is difficult to distinguish comparing to G2. Just some subjects paying attention to frequencies can point out that G4 is ``smooth", such as G3 but less than G3. This observation helps explain why G2 and G4 are challenging to differentiate in the generated group.

This also suggests that subjects tend to use the intensity of the vibrations as their primary or initial criterion for the judgment. G6 vibrations are typically composed of low-frequency components with uneven, strong, sudden pulses. G8 vibrations are composed of some different frequency components, also strong but not uneven as G6. Subjects who can pay attention to the frequency component can easily point out the differences but many subjects just judge from the intensity and say they are similar. This phenomenon appears in both groups. Observing the waveforms generated by users in Fig. \ref{fig:GenerationResults_wave} we notice that the model tends to produce uniform vibration patterns, making the generated vibrations similar to both G2 and G8. This similarity causes confusion during classification as well. 

Additionally, although the model incorporates GAN to mitigate detail loss owing to feature space compression, some distortion is inevitable. Consequently, the generated vibrations can exhibit an ``oversmoothing" effect similar to that observed in autoencoders, making them less ``sharp" compared to the real vibrations. This can reduce the uneven pulses present in the vibrations. Similarly, the intensity of the originally strong and rough surface, G8, is weakened. Thus, it is more similar to the more uniform and weaker G2. This aligns with subject feedback and the observation mentioned earlier that subjects tend to prioritize amplitude when making initial judgments.

Our demonstrated model can generate tactile vibrations that are distinguishable and aligned with target characteristics according to the user's preference. However, there is a gap between the classification task accuracy of real samples and generated samples. Although there are some differences between the vibration feedback and textures, vibrations can activate some related receptors and provide some similar perceptions \cite{konyo2005tactile}. Therefore, some related conclusions about texture perceptions may explain phenomenons in this research. A previous study points out that, eliminating movement constricted the subjective range of texture roughness or other surface features \cite{hollins2000evidence}. Because the haptic vibrations presented in this study are time-related rather than space-related, the participants did not experience spatial movement during their perception process. However, vibrational and spatial stimulation considered two components of texture perception \cite{katz2013world}. The accuracy of classification task may be influenced by this type of stimulation absence, especially in experiment 3 of Stage II, which focuses on the detailed differences in the same class.

Additionally, the phenomenon in which users can distinguish between different classes but struggle to differentiate samples within the same class might be owing to the model capturing more general information while having difficulty in capturing more specific information. This may be another type of mode collapse. The mode collapse can occur in one of two forms: the intra-class mode collapse and the inter-class mode collapse \cite{saad2022addressing}. Although we have incorporated conditional information to enable the model to generate samples across multiple categories and to control the generation process, thereby avoiding inter-category mode collapse, intra-category mode collapse can occur. 
This can lead to the condition occurring in Stage I, where subjects, while exploring similar feature spaces, encounter the model's tendency to produce similar results for a specific class or neighboring features. Consequently, this results in the model's inability to capture specific details accurately, leading to a lower distinction accuracy among subjects in experiment 3, Stage II.

Therefore, in future work, we plan to explore other generative models to enhance the diversity and precision of the generated vibrations. For example, incorporating diffusion models can help improve the variety and fidelity of the output \cite{ho2020denoising}. Diffusion models can be easy to train compared with GAN and can provide high quality results. Additionally, the current optimization process is time consuming and requires subjects to focus intensely on detecting subtle differences while adjusting the slider position. Streamlining the interface for more intuitive operation can significantly improve user experience. Our current presentation system is designed specifically to deliver texture-related vibratory feedback on simple devices, such as game controllers, rather than to replicate the sensation of real textures themselves. Therefore, in future research, we may further explore the application of this framework to spatial and temporal stimuli and aim to deploy it on equipment better suited to the original data. For instance, we envision using devices like the PHANToM, to fit the data collection method better and consider spatial stimuli to create more realistic tactile feedback.


\section{Conclusion}
Our study demonstrates the feasibility and effectiveness of a human-in-the-loop vibration generation system, TexSenseGAN, to create realistic and distinguishable haptic textures. In our TexSenseGAN model, we incorporated the idea of SRGAN and CAAE to generate class-controllable haptic vibrations with conditional information and encoded features. By employing DSS and the pre-trained generative model, we enable users to intuitively control high-dimensional latent space and generate vibration samples using simple, one-dimensional sliders. The user experiments conducted with this system indicate that it can produce vibration patterns that are recognizable and distinguishable by subjects, though challenges remain in capturing finer, specific characteristics within specific classes. Our approach offers a novel method for generating haptic vibrations based on user preferences, providing a foundation for future advancements in the field of haptic technology and virtual reality interactions.

\bibliographystyle{IEEEtran}
\bibliography{ref}

\end{document}